\documentclass{article}

\usepackage[preprint]{neurips_2023}

\newcommand{\eg}{\emph{e.g.}\@\xspace}


\usepackage{xspace}

\usepackage{graphicx}
\usepackage{subcaption}
\usepackage{booktabs}
\usepackage{multirow}
\usepackage{makecell}

\usepackage{amsmath}
\usepackage{amssymb}

\newcommand{\ourmethod}{ID-LoRA\xspace}

\usepackage[breaklinks,colorlinks,citecolor=blue]{hyperref}
\usepackage{url}

\begin{document}

\title{\ourmethod: Identity-Driven Audio-Video Personalization with In-Context LoRA}

\author{
  Aviad Dahan\thanks{Equal contribution} \quad
  Moran Yanuka\footnotemark[1] \quad
  Noa Kraicer \quad
  Lior Wolf \quad
  Raja Giryes \\[0.3em]
  Tel Aviv University \\[0.2em]
  \url{https://id-lora.github.io/}
}

\maketitle

\begin{abstract}

Existing video personalization methods preserve visual likeness but treat
video and audio separately.  Without access to the visual scene, audio
models cannot synchronize sounds with on-screen actions; and because
classical voice-cloning models condition only on a reference recording, a text prompt cannot redirect speaking style or acoustic environment.  Although prompt-conditioned audio models could offer such control, they lack access to the visual scene.

We propose \textbf{\ourmethod} (Identity-Driven In-Context LoRA), which
jointly generates a subject's appearance and voice in a single model,
letting a text prompt, a reference image, and a short audio clip govern
both modalities together.  \ourmethod adapts the LTX-2 joint audio-video
diffusion backbone via parameter-efficient In-Context LoRA and, to our
knowledge, is the first method to personalize visual appearance and voice
within a single generative pass. Two challenges arise from this formulation.
Reference and generation tokens share the same positional-encoding
space, making them hard to distinguish; we address this with
\emph{negative temporal positions}, which place reference tokens in a
disjoint region of the RoPE space while preserving their internal
temporal structure.  Furthermore, speaker characteristics tend to be
diluted during denoising; we introduce \emph{identity guidance}, a
classifier-free guidance variant that amplifies speaker-specific
features by contrasting predictions with and without the reference
signal.
In human preference studies \ourmethod is preferred over Kling 2.6 Pro,
the leading commercial unified model with voice personalization capabilities, by 73\% of annotators for voice
similarity and 65\% for speaking style.  Automatic metrics confirm these
gains: on cross-environment settings, speaker similarity improves by 24\% over
Kling, with the gap widening as reference and target conditions diverge.
A preliminary user study further suggests that joint generation provides a
useful inductive bias for physically grounded sound synthesis.  \ourmethod
achieves these results with only ${\sim}$3K training pairs on a single
GPU. 

\end{abstract}

\begin{figure}[b!]
\centering
\includegraphics[width=\textwidth]{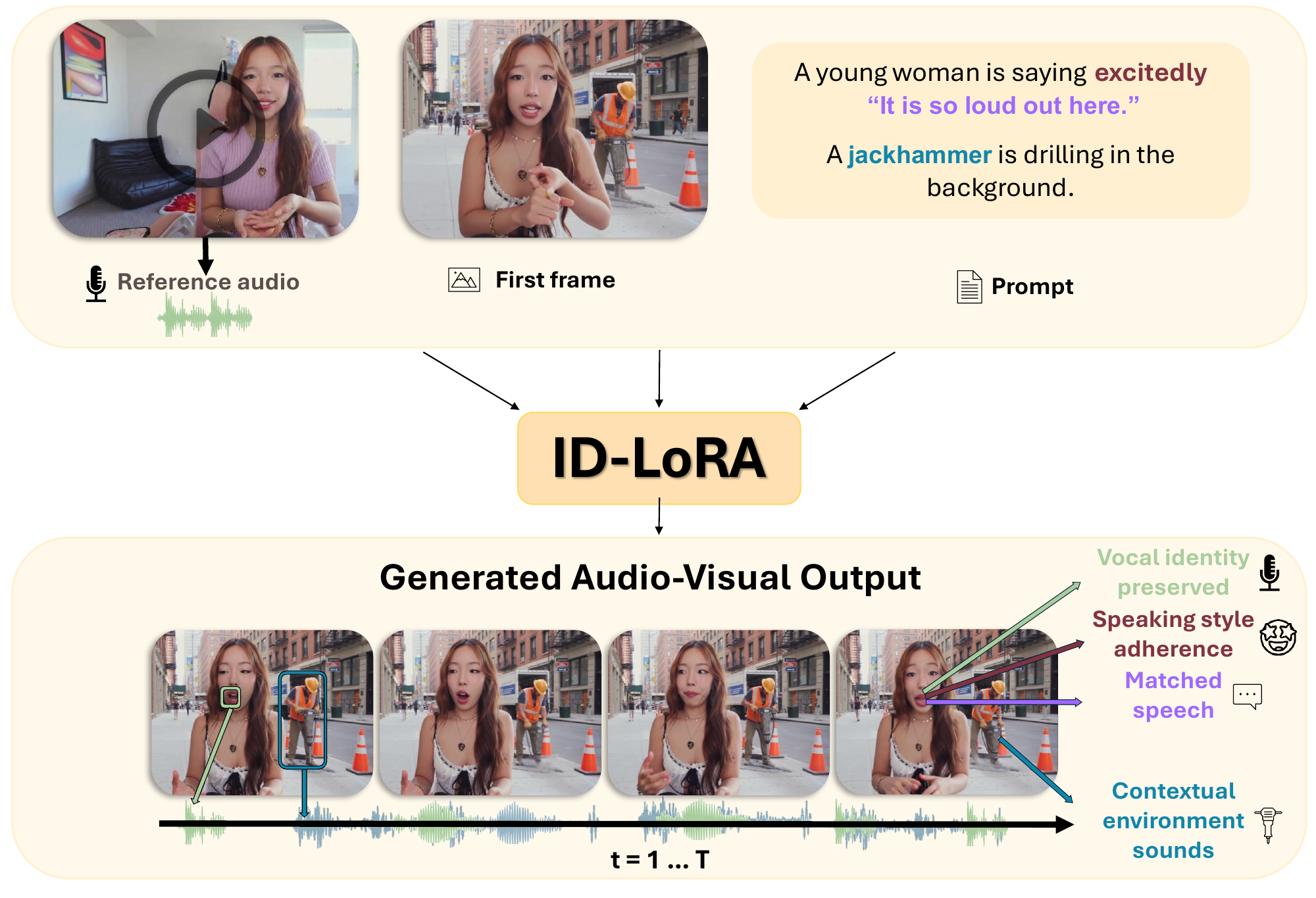}
\caption{
        \textbf{Unified Audio-Visual Personalization with \ourmethod}. \ourmethod takes a reference audio clip and a target first frame as input. Unlike cascaded pipelines that treat modalities separately, \ourmethod jointly generates synchronized video and audio in a single pass. This unified approach allows a text prompt to simultaneously dictate novel content, such as a specific speaking style and environmental acoustics (\eg, ``a jackhammer is drilling in the background''), while ensuring the subject's vocal identity and visual likeness are preserved across the generated sequence. Video examples with audio 
are available in the supplementary material.}
\label{fig:teaser}
\end{figure}

\newpage
\section{Introduction}
\label{sec:intro}

The field of generative media has witnessed a paradigm shift, moving from static image synthesis to the creation of high-fidelity, temporally coherent video. Indeed, foundational text-to-video (T2V) models, such as the open-source Wan series~\cite{wan2025wanopenadvancedlargescale}, demonstrate remarkable capabilities in generating complex scenes, and cross-modal personalization frameworks like Phantom~\cite{liu2025phantom} enable users to inject specific subject identities into these visual priors with high precision. However, human identity is inherently multimodal: a person is defined not only by their appearance but distinctly by the timbre, cadence, and prosody of their voice. Despite the visual fidelity of modern personalization methods, the resulting content remains predominantly silent, or relies on disjointed audio solutions that fracture the illusion of reality.
Current approaches to audio-visual personalization typically rely on \textit{cascaded pipelines}, where video generation is strictly conditioned on previously synthesized audio (\eg, SadTalker~\cite{zhang2023sadtalker}, VASA-1~\cite{xu2024vasa}, Hallo~\cite{xu2024hallo}). These modular approaches share a fundamental limitation: the voice-cloning stage conditions only on an audio reference and a transcript, ignoring the text prompt that describes the target scene. As a result, if a prompt requests an angry shout in a windy outdoor setting but the reference audio was recorded in a quiet studio, a cascaded pipeline will simply propagate the studio-like acoustic characteristics and neutral speaking style, failing to follow the prompt's intent. More broadly, cascaded generation prevents the prompt from jointly influencing both audio and video properties, limiting controllability over environment sounds and speaking style. While recent editing techniques like 
EditYourself \cite{flynn2026edityourself} and 
Just-Dub-It~\cite{chen2026just} have moved toward unified generation, they are fundamentally constrained to \textit{same-video} settings. They edit an existing video, maintaining the original speaker settings and acoustic environment, and cannot generalize to the \textit{cross-video} settings required to synthesize a subject in a novel context.

In this work, we address \textit{unified audio-visual personalization} (Fig.~\ref{fig:teaser}): generating video clips of a specific identity where appearance and voice are synthesized jointly within a shared latent space.  Unlike dubbing or editing approaches, our goal is to generalize to entirely new scenes. By operating within a unified latent space, we aim to place a specific subject into entirely \textit{novel} contexts while preserving both their visual likeness and vocal identity. Because the modalities are coupled, the text prompt can simultaneously dictate the scene's visual generation, environmental acoustics, and speaking style.

To achieve this, we propose \textbf{\ourmethod} (Identity-Driven In-Context LoRA), a method that adapts a unified audio-video diffusion backbone (LTX-2~\cite{hacohen2026ltx2efficientjointaudiovisual}) for audio-visual personalization. By leveraging a shared DiT backbone that processes audio and video latents, our model mimics the joint distribution of a speaker's appearance and acoustic signature, enabling identity-preserving generation in novel contexts. Crucially, we generalize the In-Context LoRA approach to allow the model to attend to the reference identity sample (a single first-frame image and a short audio clip) and generate coherent outputs in novel settings. Because both modalities share the same text conditioning, the generated audio can follow prompt descriptions of the environment and speaking style rather than being locked to the reference clip's acoustics, while simultaneously attending to the video modality.

Our contributions can be summarized as follows:
\begin{itemize}
    \item We introduce \textbf{\ourmethod}, the first In-Context LoRA framework for zero-shot joint audio-video personalization.
    \item We propose two architectural components that are key to effective identity transfer: \textit{negative temporal positions} that place reference audio tokens in a negative region of the RoPE positional space, cleanly separating them from target tokens while keeping the target at the pretrained positions; and \textit{identity guidance}, a classifier-free guidance generalization applied to the audio stream that amplifies speaker-specific features at inference time.
    \item We propose an evaluation protocol with curated splits that test different aspects of audio-visual personalization across multiple environments. We complement automatic metrics with two human evaluation paradigms: an A/B preference study and a Mean Opinion Score (MOS) study evaluating environment sound interaction across diverse physical scenarios.
\end{itemize}
We demonstrate through extensive experiments and human evaluations that our unified approach outperforms state-of-the-art cascaded baselines and commercial models in speaker similarity, lip synchronization, and audio prompt adherence.

\section{Related Work}

\subsection{Joint Audio-Video Generation}

Recent work has explored shared diffusion or flow-matching backbones that
jointly model video and audio latents.
AV-DiT~\cite{wang2024avditefficientaudiovisualdiffusion} introduces an
efficient dual-stream transformer for synchronized audio-video synthesis.
JavisDiT~\cite{liu2025javisdit} extends this paradigm to longer sequences
with improved cross-modal attention, and
Apollo~\cite{wang2026apollounifiedmultitaskaudiovideo} unifies multiple
audio-visual tasks, generation, editing, and inpainting, under a single
multi-task objective.
Ovi~\cite{low2025ovitwinbackbonecrossmodal} adopts a twin-backbone architecture with a dedicated 5B
audio branch mirroring the visual backbone of Wan2.2.
LTX-2~\cite{hacohen2026ltx2efficientjointaudiovisual}, which serves as our
backbone, employs an asymmetric 48-layer DiT with bidirectional cross-modal
attention, achieving state-of-the-art joint generation quality.
While these models produce temporally coherent audio-visual content, they are designed for general text-to-audio-video synthesis and lack mechanisms for multimodal identity personalization.

\subsection{Identity Personalization in Video and Audio}

\paragraph{Visual Identity Personalization.} 
Subject-driven image generation has advanced rapidly through
optimization-based methods such as Textual
Inversion~\cite{gal2022textual} and DreamBooth~\cite{ruiz2023dreambooth},
and through encoder-based approaches such as
IP-Adapter~\cite{ye2023ip} and InstantID~\cite{wang2024instantid} that
inject identity features into diffusion models.
Parameter-efficient fine-tuning via LoRA~\cite{hu2022lora} has become a
dominant paradigm, and In-Context LoRA
(IC-LoRA)~\cite{huang2024context} extends it by concatenating reference
and target latents along the sequence dimension, enabling zero-shot
identity transfer through self-attention without per-subject
optimization.
Video personalization builds on these foundations:
VACE~\cite{jiang2025vaceallinonevideocreation} provides an all-in-one
video creation framework, and Phantom~\cite{liu2025phantom} enables
cross-modal subject insertion with high fidelity.
Our work elevates the IC-LoRA paradigm from a visual-only technique to a comprehensive audio-visual framework. Because cross-modal identity transfer introduces severe positional entanglement during denoising, we formulate negative temporal positions and propose identity guidance, successfully handling the unique challenges of cross-modal personalization.

\paragraph{Audio Identity and Scene Control.}
Zero-shot voice cloning has progressed from speaker-embedding-conditioned
TTS to neural codec language models
~\cite{wang2023neuralcodeclanguagemodels, peng-etal-2024-voicecraft} and flow-based approaches~\cite{wang2024maskgct, du2024cosyvoice})
that achieve high-fidelity speaker transfer.
However, these models generate clean speech with no control over
background sounds or acoustic environment.
Text-to-audio models such as AudioLDM\,2~\cite{10530074} produce rich
environmental soundscapes from free-text descriptions but cannot
preserve a specific speaker's identity.
Audiobox~\cite{vyas2023audiobox} bridges this gap by combining voice
cloning with text-described background sounds, yet it operates purely in
the audio domain without access to the visual scene.
Our unified approach goes further: by generating audio and video jointly,
a single text prompt controls visual content, environment sounds, and
speaking style simultaneously, while the visual scene provides implicit
conditioning for acoustic properties such as timing and reverberation. Throughout, the speaker's vocal identity is preserved via the reference audio.

\paragraph{Audio-Driven Talking Heads.}
A related line of work generates talking-head video conditioned on a
driving audio signal~\cite{xu2024vasa, xu2024hallo, zhang2023sadtalker}.
These methods animate a source portrait to match input speech, producing
lip-synchronized results, but they require \emph{pre-existing} audio
rather than generating it, and typically operate on tightly cropped face
regions without modeling the broader scene or environment sounds.
Our setting is fundamentally different: we jointly \emph{generate} both
the audio and the video of a subject in a novel, prompt-specified context.

\subsection{Concurrent Work}
\paragraph{Audio-Visual Editing and Dubbing}
Just-Dub-It~\cite{chen2026just} adapts LTX-2 for transcript-driven video dubbing, EditYourself~\cite{flynn2026edityourself} enables transcript-based editing by inpainting the lip region, and Sync-LoRA~\cite{polaczek2025incontextsyncloraportraitvideo} propagates visual appearance changes across a video but remains video-only with no audio branch. A common thread is the \textit{same-video} constraint: all three methods modify a source video in place, inheriting its acoustic environment and speaker configuration, and none address \textit{cross-video} personalization where the subject must be synthesized in an entirely new context with potentially different environment sounds and speaking style. \ourmethod fills this gap by performing fully generative audio-visual personalization from only a reference audio clip, a first-frame image and a text prompt.

\paragraph{Identity-Aware Audio-Visual Synthesis}
DreamID-Omni~\cite{guo2026dreamidomniunifiedframeworkcontrollable} introduces a controllable framework for human-centric generation including multi-person scenarios, but requires ${\sim}$1M training samples and substantial architectural modifications. MM-Sonate~\cite{qiang2026mm} trains on ${\sim}$100M pairs across generation, editing, and animation tasks. In contrast, \ourmethod demonstrates that parameter-efficient adaptation with only ${\sim}$3K pairs can achieve strong audio-visual personalization, with particular emphasis on enabling the text prompt to govern environment sounds and speaking style.

\begin{figure}[t]
\centering
\includegraphics[width=\textwidth]{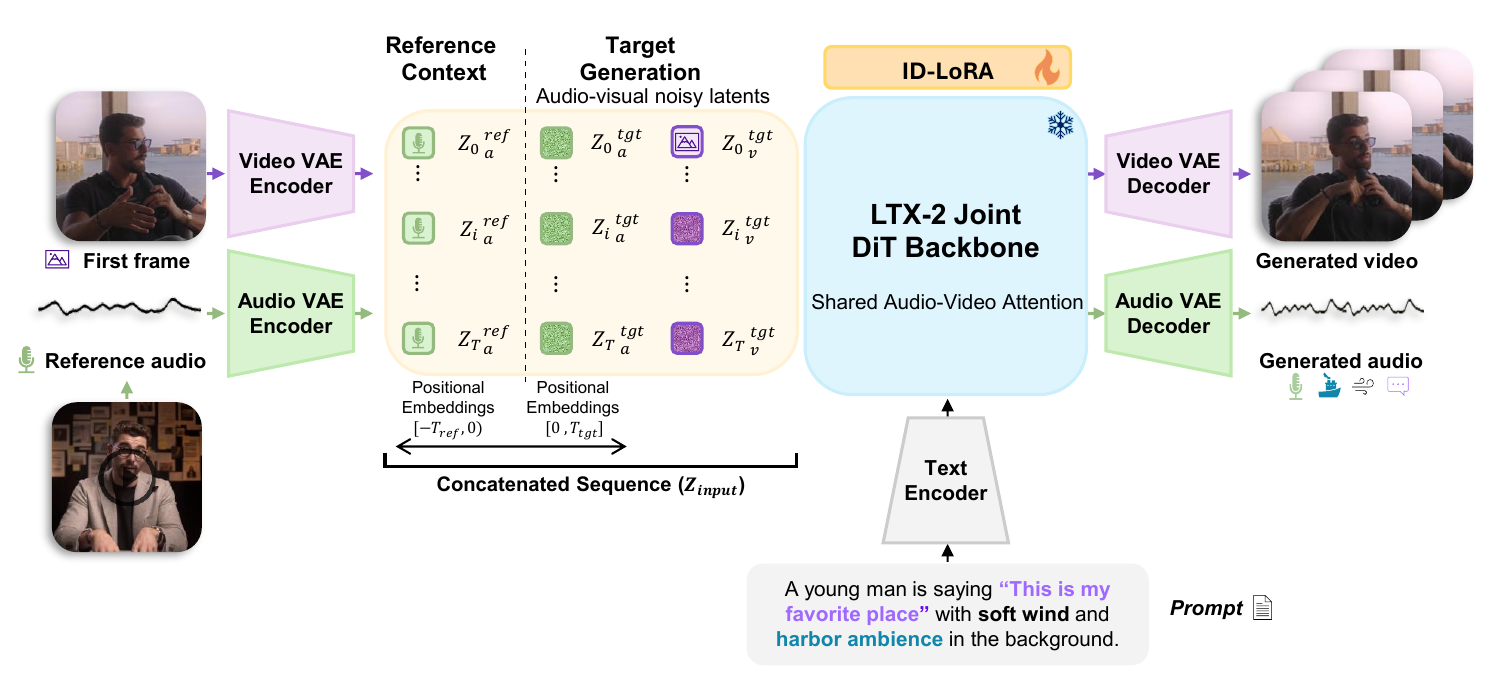}
\caption{
\textbf{\ourmethod (overview).}
Target first frame and reference audio are encoded into latents, concatenated with noisy targets, and fed to a shared LTX-2 DiT adapted with In-Context LoRA. Reference audio tokens receive \emph{negative} temporal positions in the RoPE space, cleanly separating them from the target tokens. Joint text conditioning yields synchronized, identity-preserving video and audio whose acoustics follow the prompt and adapted to the new environment.
}
\label{fig:pipeline}
\end{figure}

\section{Method}
\label{sec:method}

Our goal is to generate audio-video content that faithfully preserves both the visual appearance and vocal identity of a target individual, given only a short reference audio and a first frame. As illustrated in Fig.~\ref{fig:pipeline}, we achieve this through \textbf{\ourmethod} (Identity-Driven In-Context LoRA), which adapts a pre-trained joint audio-video diffusion backbone to perform identity transfer within a unified latent space.

\subsection{Preliminaries: Joint Audio-Video Diffusion}

We build upon LTX-2~\cite{hacohen2026ltx2efficientjointaudiovisual}, a joint audio-video generation model based on the Diffusion Transformer (DiT) architecture. LTX-2 employs an asymmetric dual-stream transformer with 48 layers that jointly processes video and audio latents through bidirectional cross-modal attention. The video stream (14B parameters) handles spatiotemporal dynamics using 3D RoPE positional encoding, while the audio stream (5B parameters) processes temporal audio features using 1D RoPE.

Both modalities are encoded into latent spaces: video pixels are compressed via a Video VAE ($32\times$ spatial, $8\times$ temporal compression), and audio waveforms are encoded through an Audio VAE that operates on mel spectrograms. 

\subsection{Identity-Driven In-Context LoRA}

In-Context LoRA (IC-LoRA)~\cite{huang2024context} enables zero-shot adaptation by concatenating reference and target latents along the sequence dimension, allowing the model to learn identity correspondence through self-attention. We generalize this paradigm to the joint audio-video setting.

\paragraph{Reference Conditioning.} Given a reference audio clip of the target speaker, we encode it into audio latents:
\begin{equation}
\mathbf{z}_a^{\text{ref}} = \mathcal{E}_a(A^{\text{ref}})
\end{equation}
where $\mathcal{E}_a$ denotes the audio VAE encoder. The reference audio latents are concatenated with the target audio latents along the sequence dimension, while the video stream uses standard text-to-video generation with first-frame conditioning, which provides a strong visual anchor for face identity while allowing temporally coherent motion and synchronized audio:
\begin{equation}
\mathbf{z}^{\text{input}} = [\mathbf{z}_v^{\text{target}}; \mathbf{z}_a^{\text{ref}}; \mathbf{z}_a^{\text{target}}]
\end{equation}
This audio-only in-context strategy allows the model to learn speaker identity transfer from the reference audio, while the video stream remains free to generate visuals guided by the text prompt and first frame while maintaining coherence with the audio.

\paragraph{Negative Temporal Positions.} A key challenge in IC-LoRA is distinguishing reference tokens from target tokens during attention. In same-video settings such as dubbing or motion transfer, prior work assigns context tokens positional encodings identical to their target counterparts, enforcing strict spatiotemporal alignment between input and output~\cite{chen2026just,polaczek2025incontextsyncloraportraitvideo}. However, in our cross-video personalization setting, the reference audio originates from a different clip entirely and shares no temporal correspondence with the target. We therefore address this by assigning \textit{negative} temporal positions to the reference audio tokens while keeping target positions positive. This creates a clear separation in the positional encoding space:
\begin{equation}
t^{\text{ref}} \in [-T_{ref}, 0), \quad t^{\text{target}} \in [0, T_{target}]
\end{equation}
where $T$ is the sequence length of the reference and target audio, respectively. This approach preserves the relative temporal structure within the reference while clearly demarcating reference from target.

\subsection{Training Objective}

We train \ourmethod using the standard diffusion denoising objective. Given a reference audio clip and a target video-audio clip from the same speaker, the model learns to denoise the target latents while attending to the reference:
\begin{equation}
\mathcal{L} = \mathbb{E}_{t, \epsilon}\left[\|\epsilon - \epsilon_\theta(\mathbf{z}_t^{\text{input}}, t, \mathbf{c})\|_2^2\right]
\end{equation}
where $\epsilon_\theta$ is the denoising network with LoRA parameters, and $\mathbf{c}$ includes text conditioning describing the target scene.


\subsection{Identity-Guided Inference}

At inference time, we introduce \textbf{Identity Guidance}, a classifier-free guidance variant that steers generation toward better identity preservation. We compute two forward passes: one with reference conditioning and one without, then extrapolate:
\begin{equation}
\hat{\epsilon} = \epsilon_\theta^{\text{uncond}} + s_{\text{id}} \cdot (\epsilon_\theta^{\text{ref}} - \epsilon_\theta^{\text{uncond}})
\end{equation}
where $s_{\text{id}}$ is the identity guidance scale. Identity guidance follows the same principle as classifier-free guidance, but applied to the reference audio rather than the text prompt. Standard CFG extrapolates away from the unconditional prediction to amplify text influence; identity guidance similarly extrapolates between the unconditional and reference-conditioned predictions, amplifying identity-specific features, vocal timbre, speaking rhythm, pronunciation, while leaving scene content and environment sounds to be governed by the text prompt.

\section{Experimental Settings}
\label{sec:experiments}

\subsection{Datasets}

We train \ourmethod on CelebV-HQ~\cite{zhu2022celebv} and TalkVid~\cite{chen2025talkvid}, maintaining separate checkpoints for each dataset. Both datasets undergo a shared preprocessing pipeline consisting of segmentation, captioning, speaker clustering, and source separation of reference audio to avoid environment leakage. For automatic evaluation, we manually curate a test set of 120 video pairs drawn from 63 held-out speakers across both datasets. The CelebV-HQ portion is divided into an \textit{easy} (same-video) split of 36 videos spanning 14 speakers and a \textit{hard} (cross-video) split of 35 videos spanning 8 speakers. The TalkVid portion comprises 49 videos from 41 held-out speakers. All captions manually verified. To assess cross-dataset generalization, we also evaluate the CelebV-HQ checkpoint on the TalkVid test set (see Sec.~\ref{sec:talkvid}). Full preprocessing and split details are provided in the supplementary material.

\subsection{Baselines}

We compare \ourmethod against three cascaded pipelines that pair a zero-shot voice-cloning model (CosyVoice 3.0~\cite{du2025cosyvoice}, VoiceCraft~\cite{peng-etal-2024-voicecraft}, or ElevenLabs~\cite{elevenlabs2026}) with WAN2.2~\cite{wan2025wanopenadvancedlargescale} as the video backbone. ElevenLabs supports built-in emotion and style control via automatic prompt enhancement, making it the strongest cascaded baseline for speaking style adherence. We also compare against Kling 2.6 Pro~\cite{kling2026}, a closed-source commercial model that generates talking-face video with voice cloning in a unified pipeline. For both ElevenLabs and Kling, prompts were adapted following each platform's official guidelines. Per-baseline details are provided in the supplementary material.

\subsection{Evaluation Metrics}

We evaluate five axes against the GT target videos: \textbf{Speaker Similarity} via WavLM+ECAPA-TDNN~\cite{Wang2021UniSpeech} cosine similarity \textbf{Face Similarity} via ArcFace~\cite{deng2019arcface} cosine similarity (5-frame average); \textbf{Lip Synchronization} via SyncNet~\cite{chung2016out, prajwal2020lip},  LSE-D ($\downarrow$) and LSE-C ($\uparrow$); \textbf{Audio Prompt Adherence} via CLAP~\cite{elizalde2023clap} similarity between generated audio and the combined environment and speaking style prompt; and \textbf{WER} via Whisper-large-v3 \cite{radford2022whisper} transcription. Full details are provided in the supplementary material.

\subsection{Implementation Details}

We implement \ourmethod on top of the LTX-2 model using LoRA with rank 128. Training is performed on a single NVIDIA H100 GPU for 6,000 steps, using AdamW optimizer with learning rate $2 \times 10^{-4}$ and batch size of 4. First-frame conditioning is applied with probability 0.9 during training.

At inference, we use 30 denoising steps with the following guidance scales: video CFG = 3.0, audio CFG = 7.0, identity guidance = 4.0, and AV-bimodal CFG = 3.0. STG (Spatio-Temporal Guidance) is applied with scale 1.0 on block 29. Generated videos are 1024$\times$1024 resolution with 121 frames at 25fps for both CelebV-HQ and TalkVid. AV-bimodal and STG guidance have been introduced in \cite{hacohen2026ltx2efficientjointaudiovisual} and \cite{hyung2024spatiotemporal} respectively.

\section{Results}
\label{sec:results}

\subsection{Main Results on CelebV-HQ}


Table~\ref{tab:main_results_wan} presents our main comparison between \ourmethod, three cascaded baselines (CosyVoice 3.0 + WAN2.2, VoiceCraft + WAN2.2, and ElevenLabs + WAN2.2), and the state-of-the-art closed-source Kling 2.6 Pro model across all evaluation splits: CelebV-HQ easy (same-video), CelebV-HQ hard (cross-video), and TalkVid. All cascaded baselines use WAN2.2~\cite{wan2025wanopenadvancedlargescale} as the video backbone. We additionally isolate architecture effects by comparing against LTX-based variants in Table~\ref{tab:main_results_ltx_variants}.

\begin{table}[t]
    \centering
    \resizebox{\textwidth}{!}{%
    \begin{tabular}{@{}clcccccc@{}}
    \toprule
    & Method  & Spk Sim $\uparrow$ & Face Sim $\uparrow$ & LSE-D $\downarrow$ & LSE-C $\uparrow$ & CLAP $\uparrow$ & WER $\downarrow$ \\
    \midrule
    \multirow{5}{*}{\rotatebox[origin=c]{90}{\large\textbf{C-Easy}}}
    & CosyVoice 3.0 + WAN2.2  & \underline{0.510} & \textbf{0.886} & 11.27 & 1.53 & 0.301 & 0.273 \\
    & VoiceCraft + WAN2.2     & 0.412 & \underline{0.879} & 10.99 & 1.48 & \underline{0.335} & 0.241 \\
    & ElevenLabs + WAN2.2     & 0.413 & \underline{0.879} & 11.98 & 1.51 & 0.266 & 0.180 \\
    & Kling 2.6 Pro$^\dagger$       & 0.487 & 0.847 & \underline{10.24} & \underline{3.01} & 0.327 & \textbf{0.046} \\
    & \textbf{\ourmethod (Ours)} & \textbf{0.573} & 0.870 & \textbf{8.20} & \textbf{4.43} & \textbf{0.372} & \underline{0.106} \\
    \midrule
    \multirow{5}{*}{\rotatebox[origin=c]{90}{\large\textbf{C-Hard}}}
    & CosyVoice 3.0 + WAN2.2      & \underline{0.391} & 0.890 & 11.40 & 1.50 & 0.249 & 0.362 \\
    & VoiceCraft + WAN2.2     & 0.344 & \underline{0.892} & 10.60 & 1.33 & 0.258 & 0.427 \\
    & ElevenLabs + WAN2.2     & 0.357 & \textbf{0.894} & 11.86 & 1.72 & 0.238 & 0.154 \\
    & Kling 2.6 Pro$^\dagger$       & 0.385 & 0.854 & \underline{9.49} & \underline{3.47} & \underline{0.316} & \underline{0.121} \\
    & \textbf{\ourmethod (Ours)} & \textbf{0.477} & 0.874 & \textbf{8.49} & \textbf{3.90} & \textbf{0.363} & \textbf{0.113} \\
    \midrule
    \multirow{6}{*}{\rotatebox[origin=c]{90}{\large\textbf{TalkVid}}}
    & CosyVoice 3.0 + WAN2.2      & 0.579 & 0.770 & 12.34 & 1.20 & 0.315 & 0.223 \\
    & VoiceCraft + WAN2.2     & 0.457 & \textbf{0.773} & 12.15 & 1.22 & \underline{0.407} & 0.171 \\
    & ElevenLabs + WAN2.2     & 0.491 & 0.772 & 12.42 & 1.31 & 0.319 & 0.064 \\
    & Kling 2.6 Pro$^\dagger$       & 0.506 & 0.754 & 11.59 & 2.40 & 0.326 & \textbf{0.040} \\
    & \textbf{\ourmethod (Ours)} & \textbf{0.599} & \underline{0.772} & \underline{10.62} & \underline{3.09} & 0.385 & \underline{0.054} \\
    & \makecell[l]{\textbf{\ourmethod}\\\textbf{(Celeb $\rightarrow$ TalkVid)}} & \underline{0.595} & 0.767 & \textbf{10.32} & \textbf{3.12} & \textbf{0.412} & 0.065 \\
    \bottomrule
    \end{tabular}}
    \vspace{7pt}
    \caption{Comparison on CelebV-HQ easy (same-video), hard (cross-video), and TalkVid splits. $\uparrow$~higher is better, $\downarrow$~lower is  better. Best in \textbf{bold}, second best \underline{underlined}.
  $^\dagger$~Commercial model.}
    \label{tab:main_results_wan}
\end{table}

On the \textbf{easy} (same-video) subset, all methods benefit from high source-target similarity, representing a voice replication scenario. On the \textbf{hard} (cross-video) subset, which tests generalization to novel acoustic settings, our advantages become more pronounced: the speaker similarity gap over the best cascaded baseline widens from +0.063 (easy) to +0.086 (hard). This suggests that our unified approach generalizes more robustly to new settings, whereas cascaded pipelines degrade more sharply when reference and target conditions diverge.

\ourmethod outperforms all cascaded baselines on \textbf{speaker similarity} and \textbf{lip synchronization}, demonstrating that a unified generation scheme better preserves the target voice and produces more coherent audio-visual alignment than cascading separately trained models. Kling 2.6 Pro, a unified commercial solution trained on significantly larger data, does not surpass our open approach on these axes. Both unified approaches (ours and Kling) also achieve substantially higher audio prompt adherence (CLAP) than the cascaded baselines, with the gap widening on the hard split (+0.105 vs. best cascaded) where environments diverge and cascaded pipelines blindly propagate reference acoustics. Statistical significance tests confirm our speaker similarity gains over all baselines at p < 0.001, and our CLAP advantage over Kling reaches p<0.01.


\subsection{Cross-Dataset Generalization}
\label{sec:talkvid}

We additionally report TalkVid results in Table~\ref{tab:main_results_wan}. \ourmethod maintains its advantages in speaker similarity and lip synchronization on this dataset. Notably, the CelebV-HQ checkpoint applied to TalkVid without fine-tuning achieves speaker similarity of 0.595, only 0.004 below the in-domain checkpoint, still surpassing all other baselines, while actually improving CLAP (0.412 vs. 0.385) and LSE-D (10.32 vs. 10.62). This likely reflects CelebV-HQ's greater acoustic diversity, indicating that \ourmethod learns transferable identity representations rather than dataset-specific shortcuts.



\begin{table}[t]
    \centering
    \begin{tabular}{@{}clcccccc@{}}
    \toprule
    & Method  & Spk Sim $\uparrow$ & Face Sim $\uparrow$ & LSE-D $\downarrow$ & LSE-C $\uparrow$ & CLAP $\uparrow$ & WER $\downarrow$ \\
    \midrule
    \multirow{5}{*}{\rotatebox[origin=c]{90}{\large\textbf{C-Easy}}}
    & CosyVoice 3.0 + LTX         & \underline{0.522} & 0.875 & 9.42 & 2.81 & 0.307 & 0.287 \\
    & VoiceCraft + LTX        & 0.412 & \textbf{0.885} & 9.19 & 2.32 & \underline{0.335} & 0.241 \\
    & ElevenLabs + LTX        & 0.420 & 0.873 & 9.44 & \underline{2.99} & 0.262 & \underline{0.185} \\
    & LTX-Zeroshot & 0.462 & \underline{0.881} & \textbf{8.05} & 2.41 & 0.308 & 4.634 \\
    & \textbf{\ourmethod} & \textbf{0.573} & 0.870 & \underline{8.20} & \textbf{4.43} & \textbf{0.372} & \textbf{0.106} \\
    \midrule
    \multirow{5}{*}{\rotatebox[origin=c]{90}{\large\textbf{C-Hard}}}
    & CosyVoice 3.0 + LTX         & \underline{0.409} & \underline{0.888} & 9.39 & 2.13 & 0.246 & 0.362 \\
    & VoiceCraft + LTX        & 0.344 & 0.883 & 8.72 & 1.52 & \underline{0.258} & 0.427 \\
    & ElevenLabs + LTX        & 0.364 & 0.886 & 10.31 & \underline{2.21} & 0.235 & \underline{0.152} \\
    & LTX-Zeroshot & 0.374 & \textbf{0.896} & \textbf{7.89} & 2.07 & 0.242 & 2.393 \\
    & \textbf{\ourmethod} & \textbf{0.477} & 0.874 & \underline{8.49} & \textbf{3.90} & \textbf{0.363} & \textbf{0.113} \\
    \midrule
    \multirow{6}{*}{\rotatebox[origin=c]{90}{\large\textbf{TalkVid}}}
    & CosyVoice 3.0 + LTX & 0.578 & 0.768 & 11.33 & 2.33 & 0.324 & 0.223 \\
    & VoiceCraft + LTX & 0.457 & 0.768 & 11.37 & 1.81 & \underline{0.405} & 0.165 \\
    & ElevenLabs + LTX & 0.493 & 0.768 & 11.13 & 2.71 & 0.322 & \underline{0.064} \\
    & LTX-Zeroshot & 0.530 & \textbf{0.850} & \textbf{9.81} & 2.45 & 0.381 & 0.471 \\
    & \textbf{\ourmethod } & \textbf{0.599} & \underline{0.772} & 10.62 & \underline{3.09} & 0.385 & \textbf{0.054} \\
    & \makecell[l]{\textbf{\ourmethod}\\(\textbf{Celeb $\rightarrow$ TalkVid})} & \underline{0.595} & 0.767 & \underline{10.32} & \textbf{3.12} & \textbf{0.412} & 0.065 \\
    \bottomrule
    \end{tabular}
    \vspace{7pt}
    \caption{Ours vs LTX variants across three datasets: CelebV-HQ easy (same-video), CelebV-HQ hard (cross-video), and TalkVid. Includes TTS-conditioned LTX and LTX-Zeroshot variants. $\uparrow$ higher is better, $\downarrow$ lower is better. Best in \textbf{bold}, second best \underline{underlined}.}
    \label{tab:main_results_ltx_variants}
\end{table}

\subsection{Controlling for Backbone Effects}
\label{sec:ltx}
To isolate pipeline design from model family effects, Table~\ref{tab:main_results_ltx_variants} compares \ourmethod with LTX-based cascaded variants (CosyVoice-LTX, VoiceCraft-LTX, ElevenLabs-LTX) and LTX-Zeroshot.

Across CelebV-HQ easy/hard, \ourmethod remains the strongest on speaker similarity and synchronization-oriented metrics (LSE-C and competitive LSE-D), while several LTX variants achieve higher face similarity, an effect we analyze below. On TalkVid, \ourmethod still leads in speaker similarity and low WER, whereas LTX-Zeroshot can obtain very low LSE-D together with degraded transcription quality (high WER), indicating that low SyncNet distance alone can overestimate quality when articulation is poor. This motivates evaluating synchronization jointly with intelligibility and identity rather than in isolation.

\paragraph{Face similarity and lip synchronization.}
All baselines achieve comparable or higher face similarity than \ourmethod, while Kling 2.6 Pro scores lower (0.847–0.854). This reflects a systematic bias in ArcFace: methods with less lip motion keep face embeddings closer to the reference, inflating the metric. A similar confound affects lip synchronization: LTX-Zeroshot achieves the lowest LSE-D across all splits yet also the lowest LSE-C confidence and largely unintelligible speech (WER $>$ 2.3 on CelebV-HQ), indicating that its low lip-sync distance stems from near-static frames rather than accurate articulation. These observations motivate evaluating synchronization jointly with intelligibility and identity rather than in isolation, and suggest that face similarity alone is an incomplete measure for talking-head generation, as it penalizes realistic speech articulation. A quantitative lip motion analysis is provided in the supplementary material.


\subsection{Ablation Studies}

\begin{table}[t]
    \centering
    \begin{tabular}{@{}lccccc@{}}
    \toprule
    Method & Spk Sim $\uparrow$ & Face Sim $\uparrow$ & LSE-D $\downarrow$ & CLAP $\uparrow$ & WER $\downarrow$ \\
    \midrule
    No Identity Guidance & 0.438 & \textbf{0.875} & 8.71 & \textbf{0.364} & 0.113 \\
    Standard PEs & 0.441 & 0.872 & 9.02 & 0.359 & 0.252 \\
    \textbf{OURS (Neg. PEs + IdG)} & \textbf{0.477} & \underline{0.874} & \textbf{8.50} & \underline{0.363} & \textbf{0.113} \\
    \bottomrule
    \end{tabular}
    \vspace{7pt}
    \caption{Core design ablation on the CelebV-HQ hard split (cross-video): no identity guidance, standard positional encoding, and full OURS (negative positions + identity guidance). The full setup gives the best overall balance on synchronization and intelligibility while retaining strong identity.}
    \label{tab:negpos_ablation_nomask}
    \end{table}

\paragraph{Negative Temporal Positions and Identity-guidance.} Table~\ref{tab:negpos_ablation_nomask} presents a core design ablation comparing 3 configurations on the hard split. Replacing negative temporal positions with standard positional encodings degrades lip synchronization (LSE-D: 9.02 vs.\ 8.50) and sharply increases WER (0.252 vs.\ 0.113), indicating that the model struggles to separate reference from target speech without a clear positional boundary. Removing identity guidance reduces speaker similarity (0.438 vs.\ 0.477), confirming that identity guidance steers the model toward speaker-specific features. A sensitivity analysis of the identity guidance scale is provided in the supplementary material.

\begin{figure}[t]
\centering
\begin{subfigure}[t]{0.48\linewidth}
 \centering
\includegraphics[width=\linewidth]{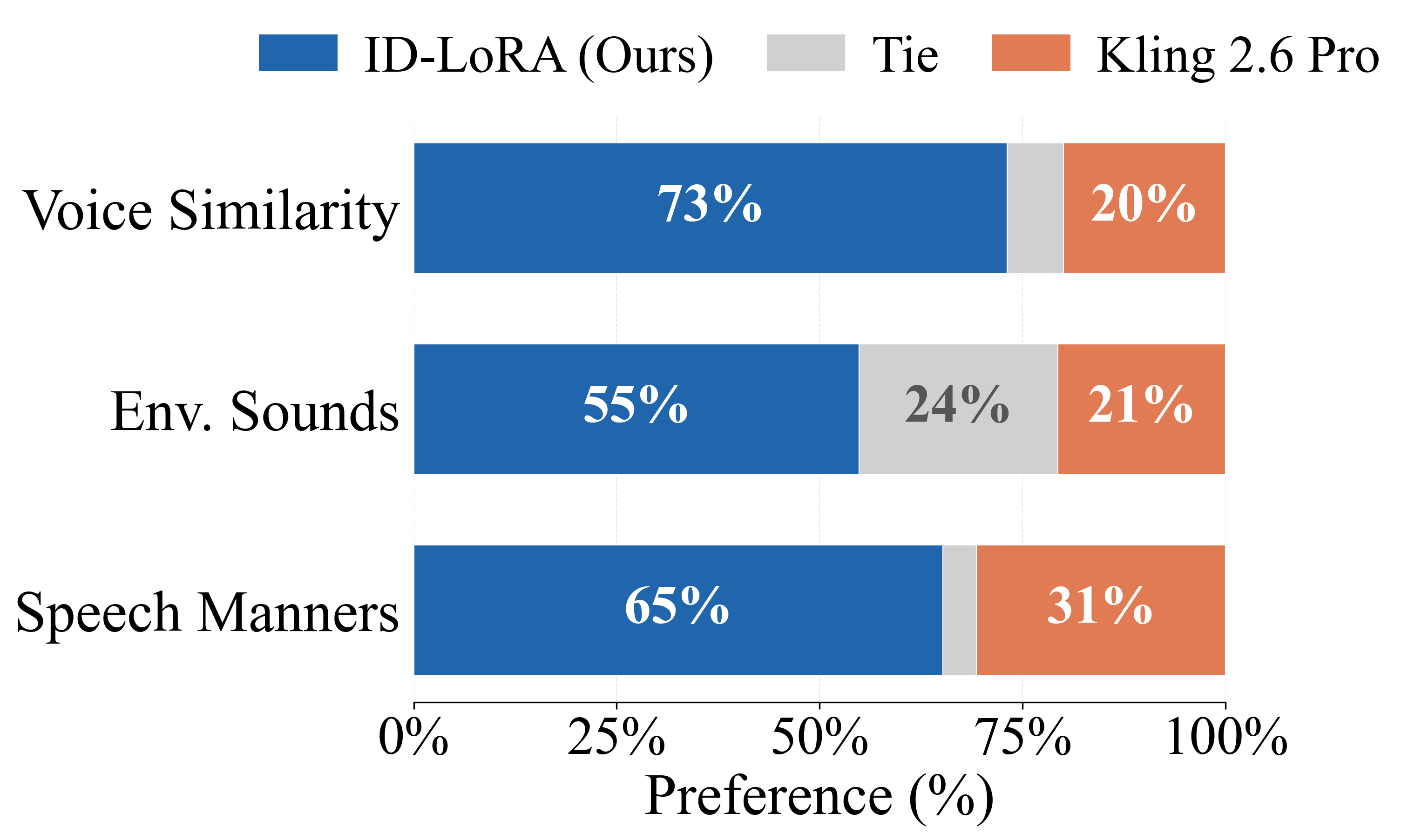}
\caption{\ourmethod vs.\ Kling 2.6 Pro}
\label{fig:user_study_kling}
\end{subfigure}
\hfill
\begin{subfigure}[t]{0.48\linewidth}
\centering
\includegraphics[width=\linewidth]{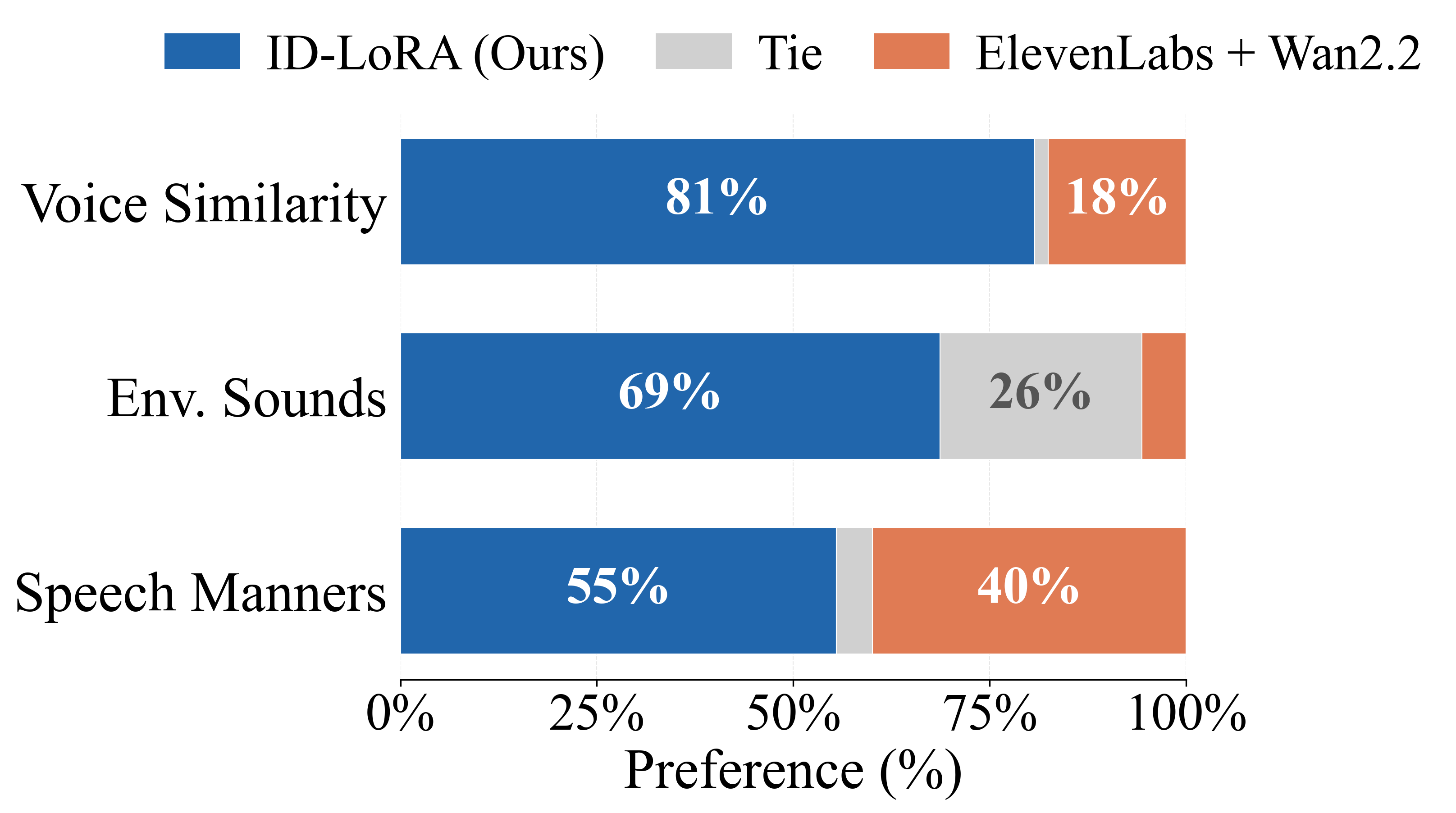}
\caption{\ourmethod vs.\ ElevenLabs + Wan 2.2}
\label{fig:user_study_elevenlabs}
\end{subfigure}
\caption{Human evaluation results: A/B preference rates (\%) on the hard (cross-video) split. Annotators on AMT evaluated 35 pairs across 8 speakers along three axes: voice similarity, environment sounds, and speech manners.}
\label{fig:user_study}
\end{figure}

\begin{figure}[t]
\centering
\setlength{\tabcolsep}{2pt}

\begin{tabular}{ccccc}

\includegraphics[width=0.18\linewidth]{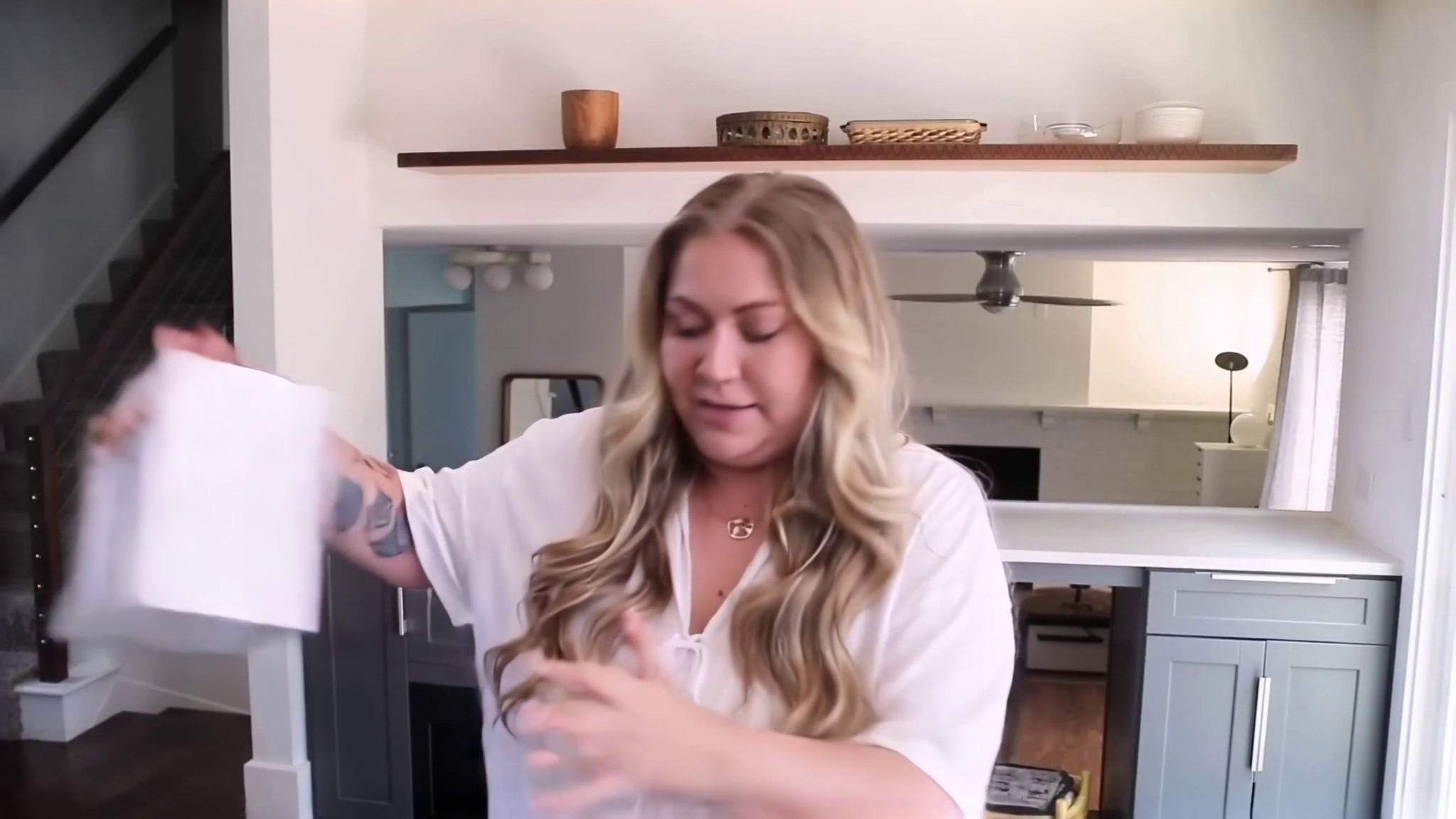} &
\includegraphics[width=0.18\linewidth]{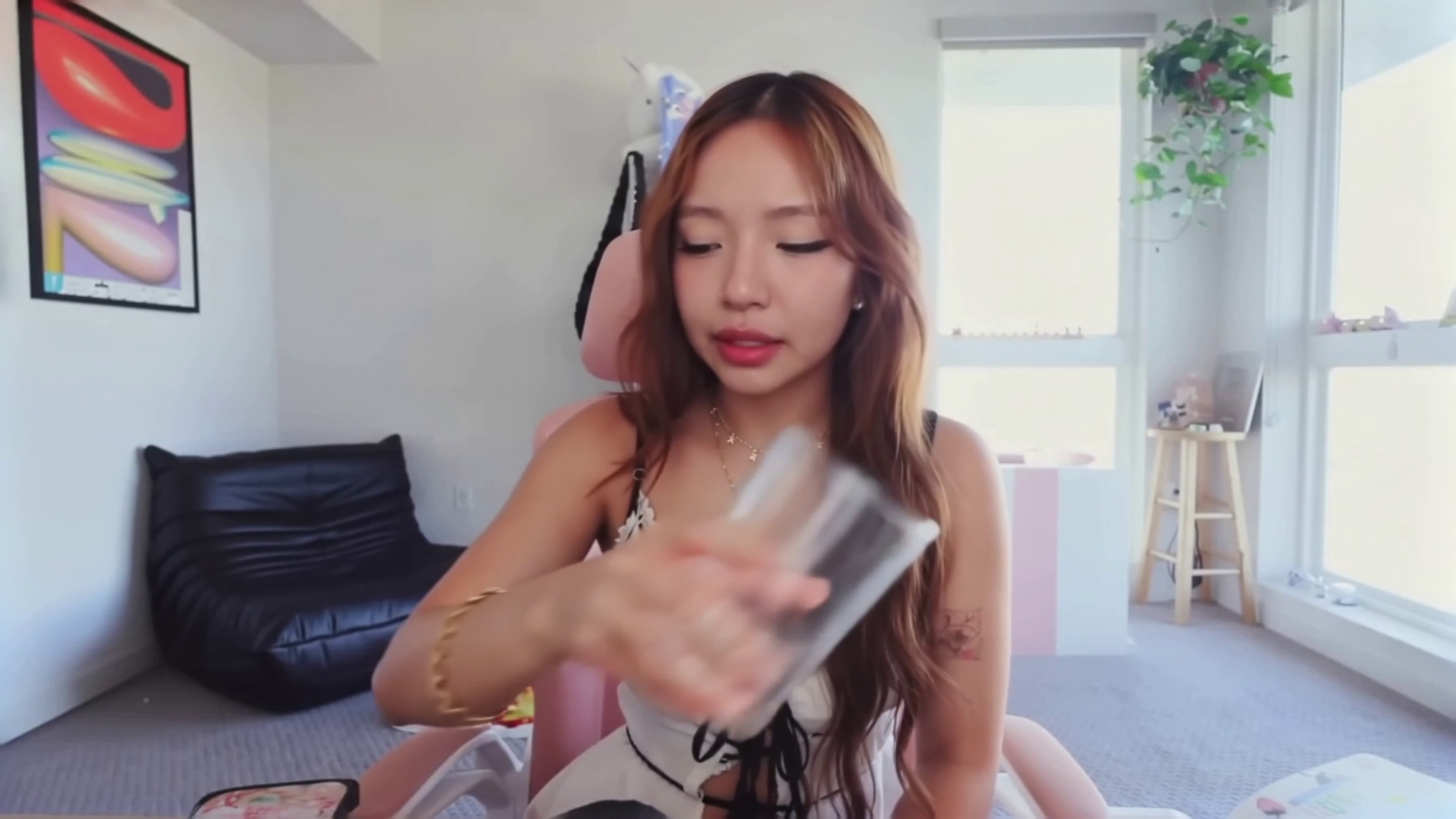} &
\includegraphics[width=0.18\linewidth]{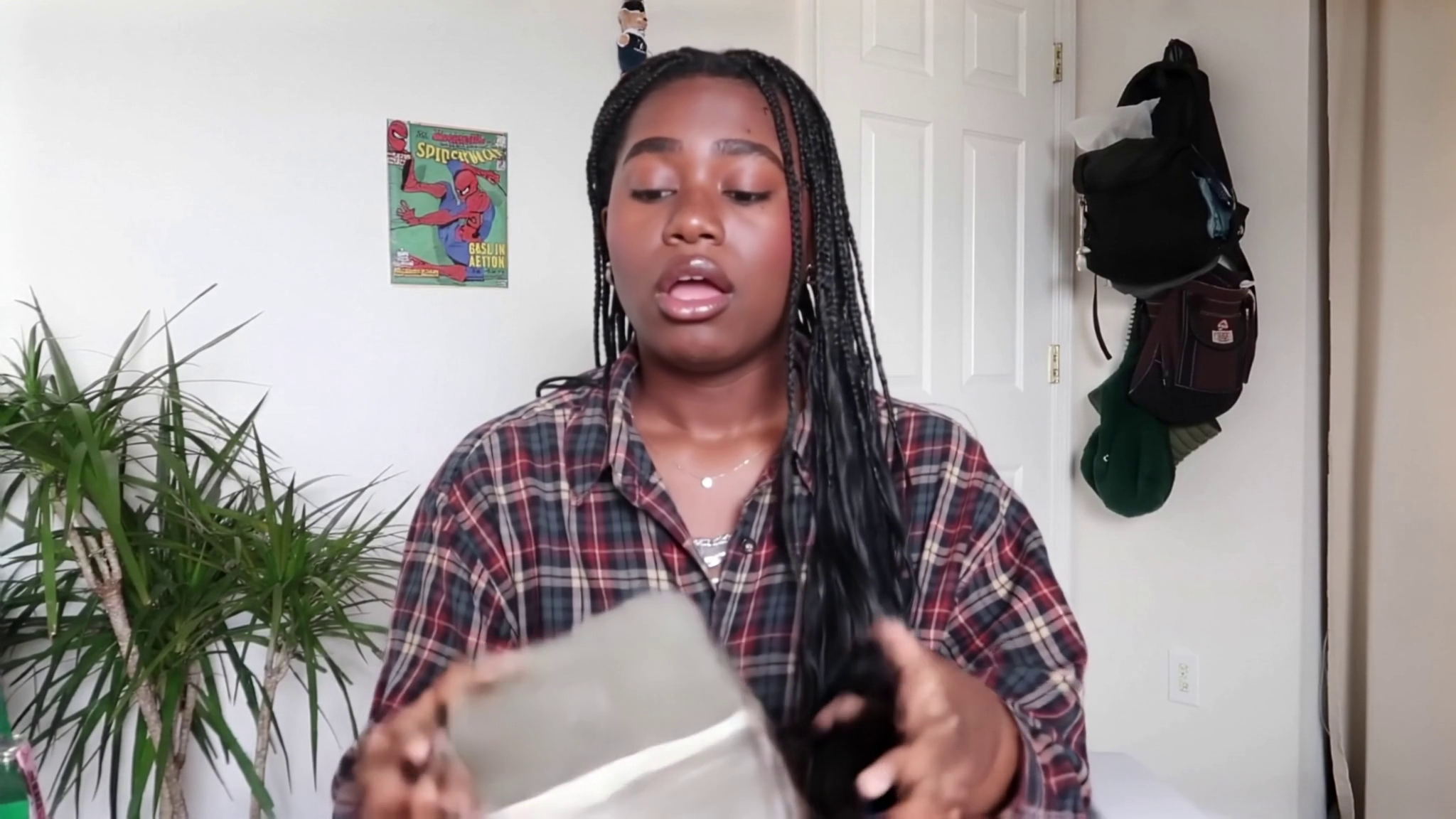} &
\includegraphics[width=0.18\linewidth]{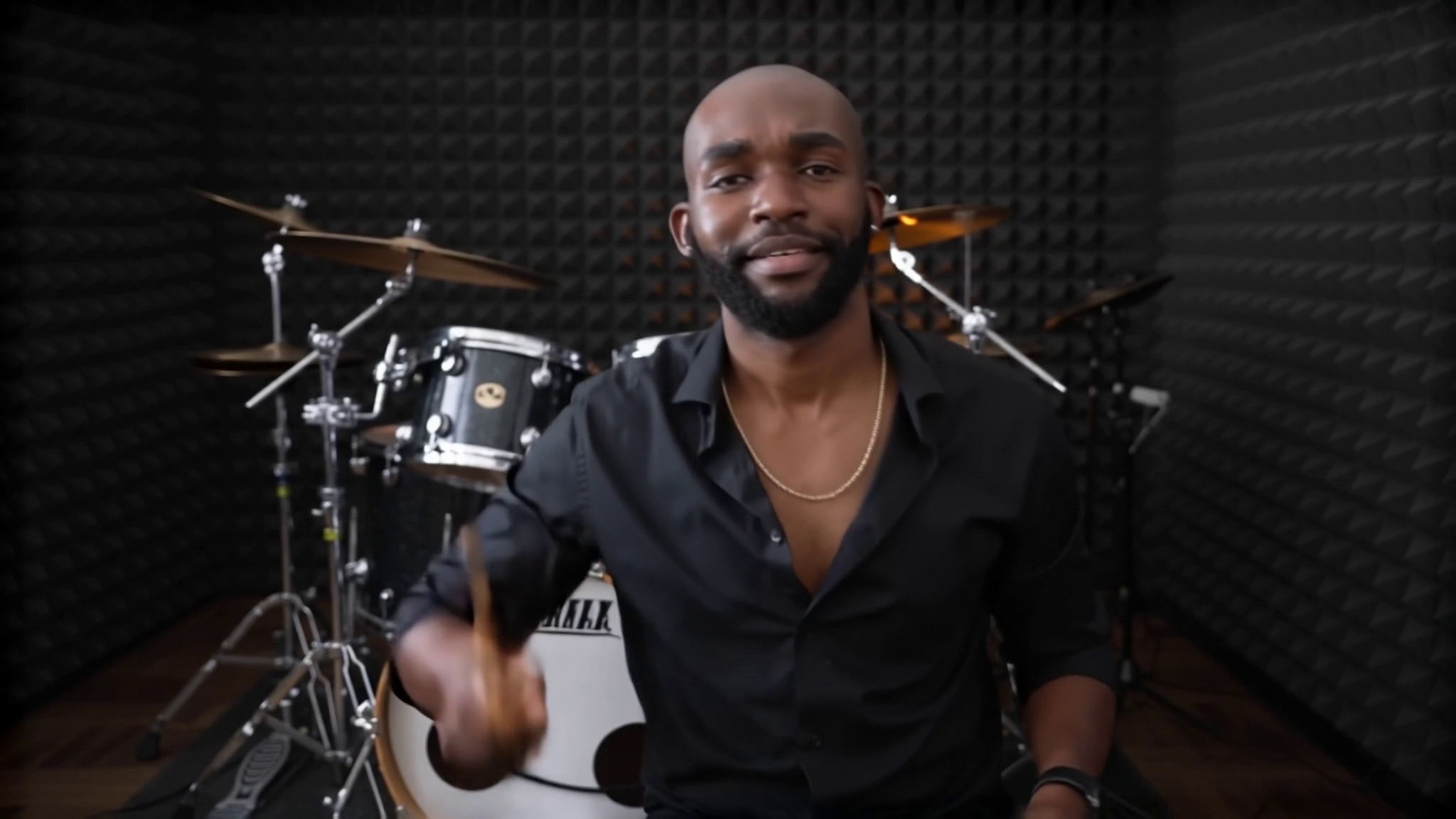} &
\includegraphics[width=0.18\linewidth]{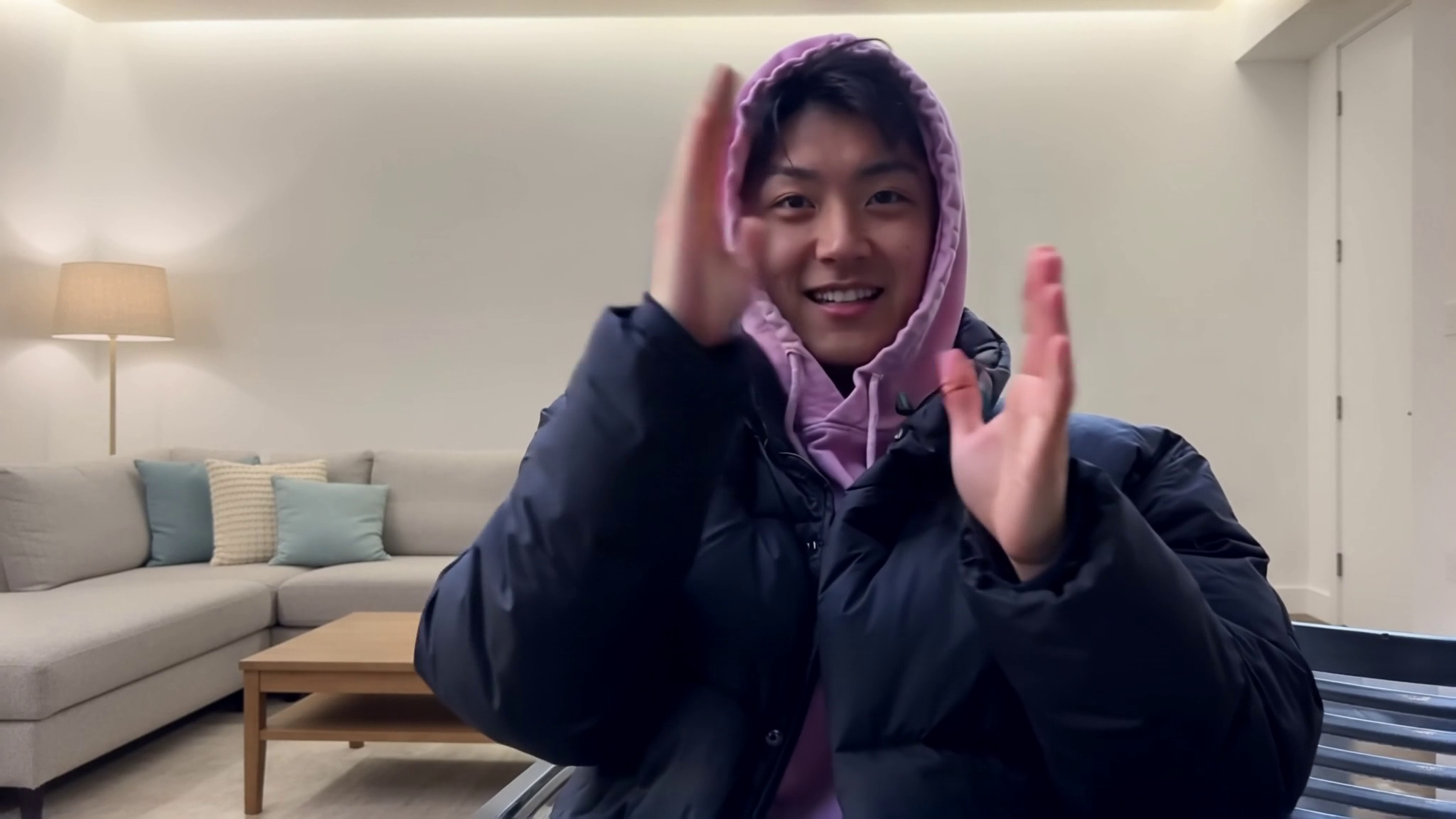} \\

{\scriptsize Box Thud} &
{\scriptsize Glass Break} &
{\scriptsize Metal Box} &
{\scriptsize Drums} &
{\scriptsize Hand Clap} \\[2pt]

\includegraphics[width=0.18\linewidth]{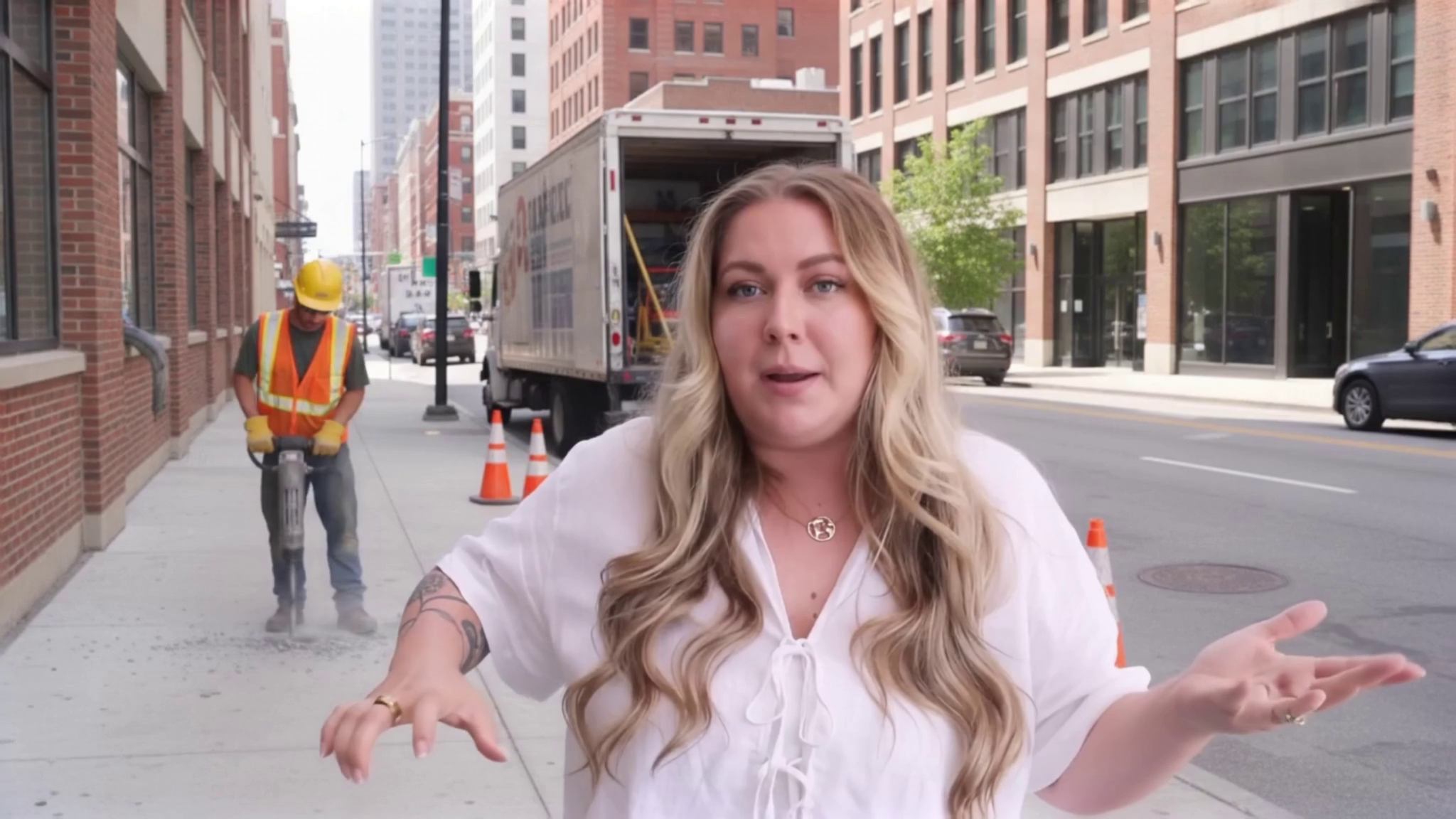} &
\includegraphics[width=0.18\linewidth]{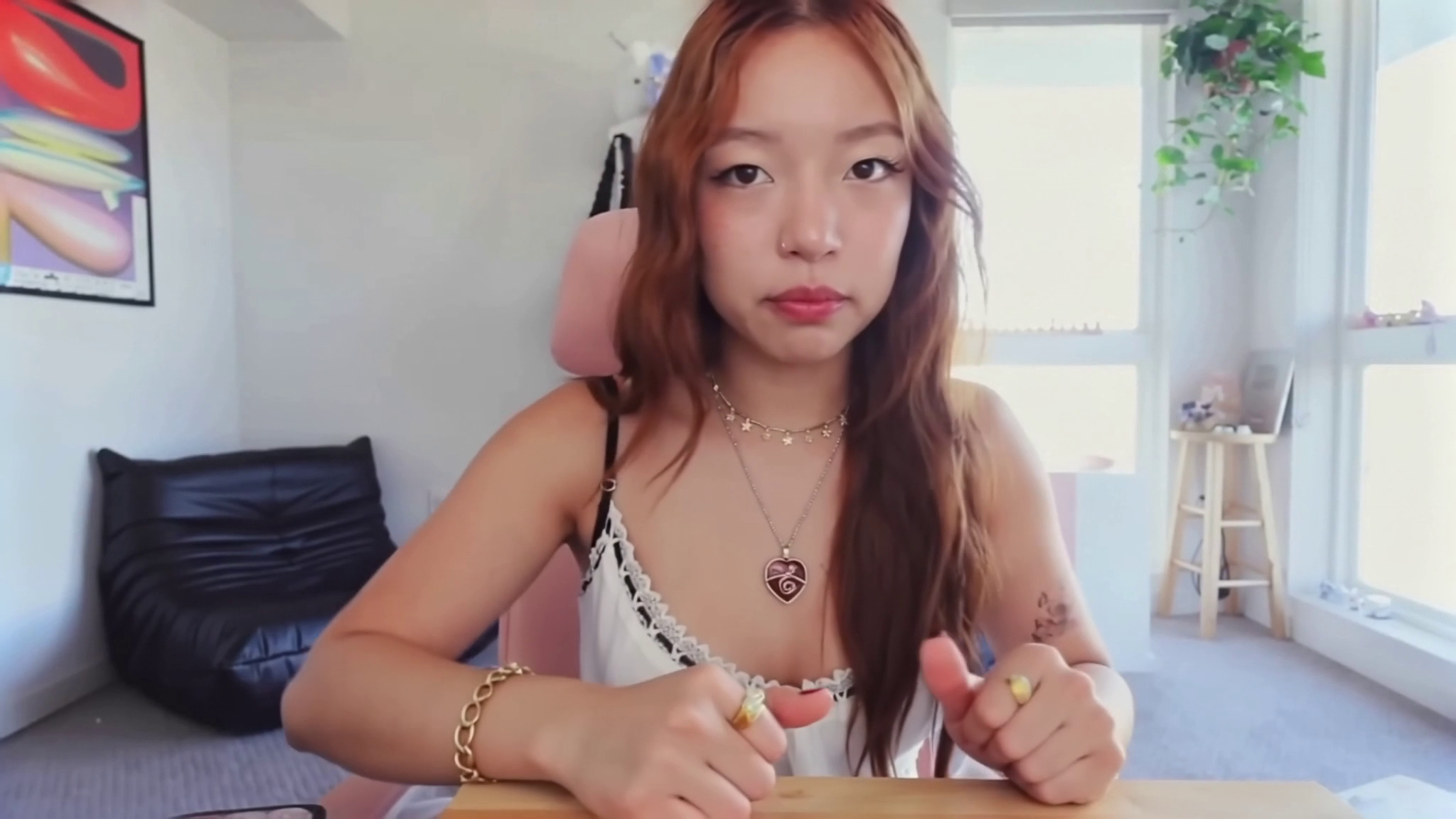} &
\includegraphics[width=0.18\linewidth]{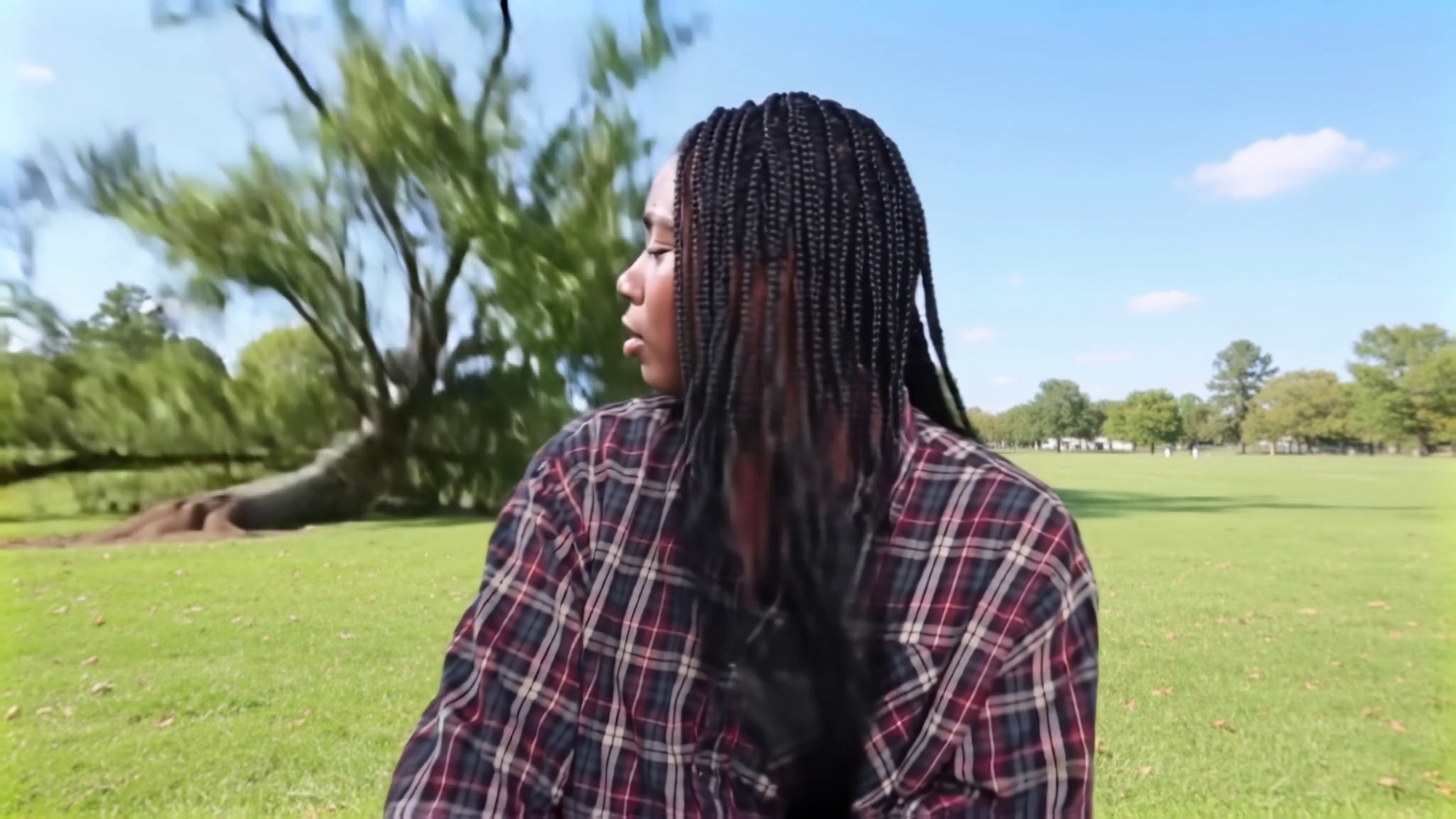} &
\includegraphics[width=0.18\linewidth]{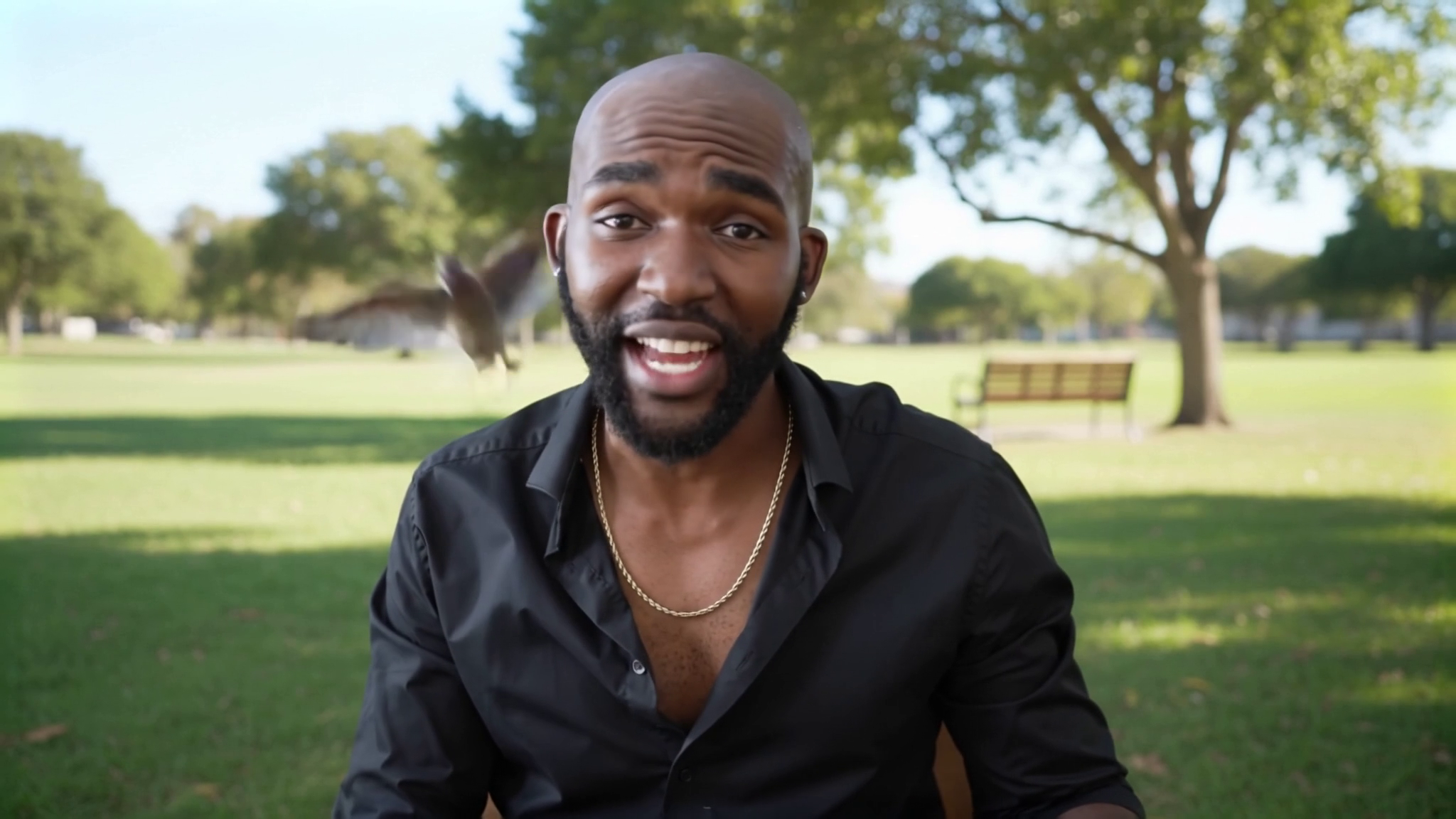} &
\includegraphics[width=0.18\linewidth]{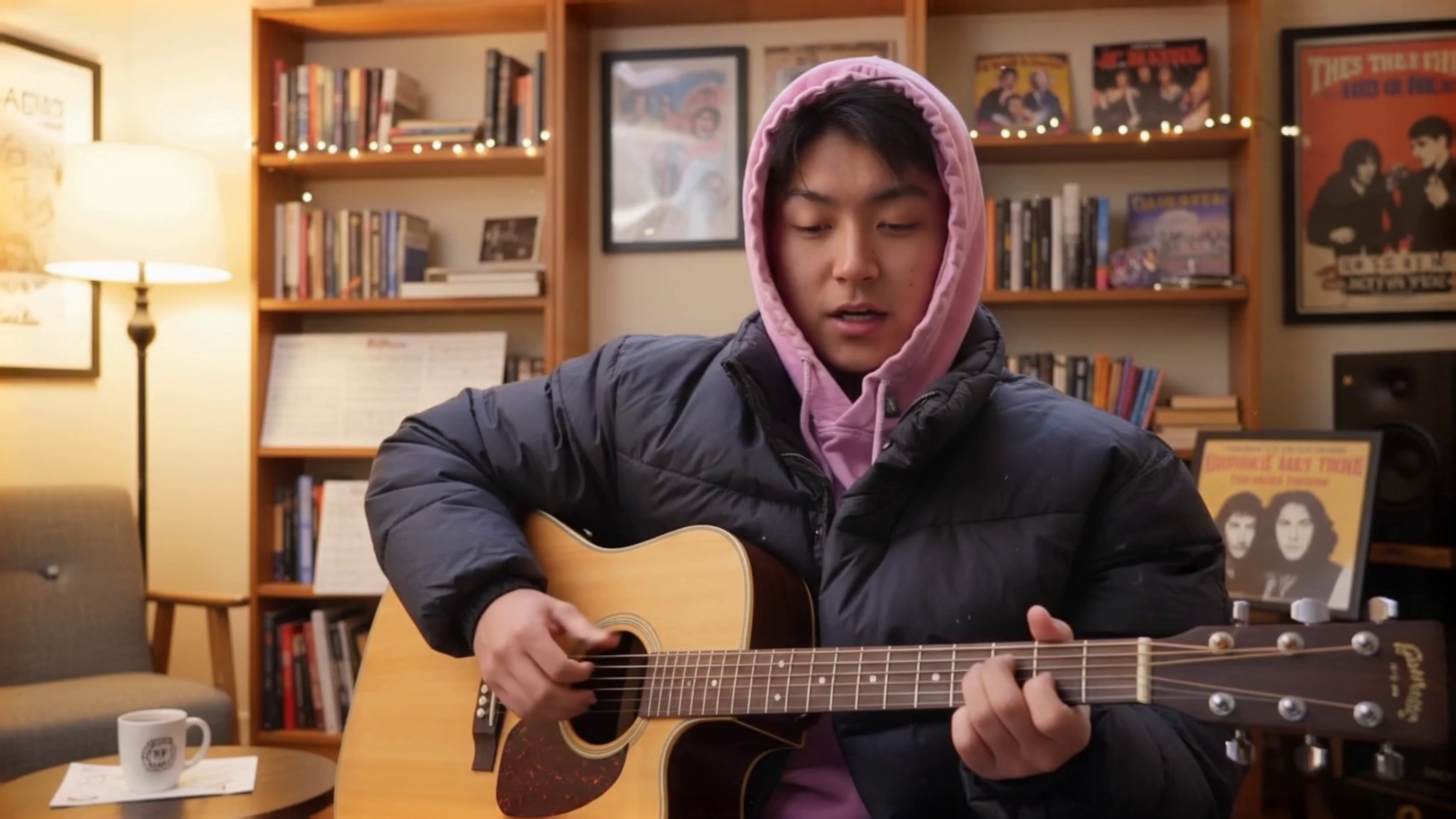} \\

{\scriptsize Jackhammer} &
{\scriptsize Knocking} &
{\scriptsize Tree Fall} &
{\scriptsize Bird Chirp} &
{\scriptsize Guitar}

\end{tabular}

\caption{Examples from the MOS evaluation set. Each column shows two generated videos from the same speaker under different environmental sound conditions.}
\label{fig:mos_examples}

\end{figure}

\subsection{Human Evaluation}

\begin{figure}[t]
\centering
\includegraphics[width=\linewidth]{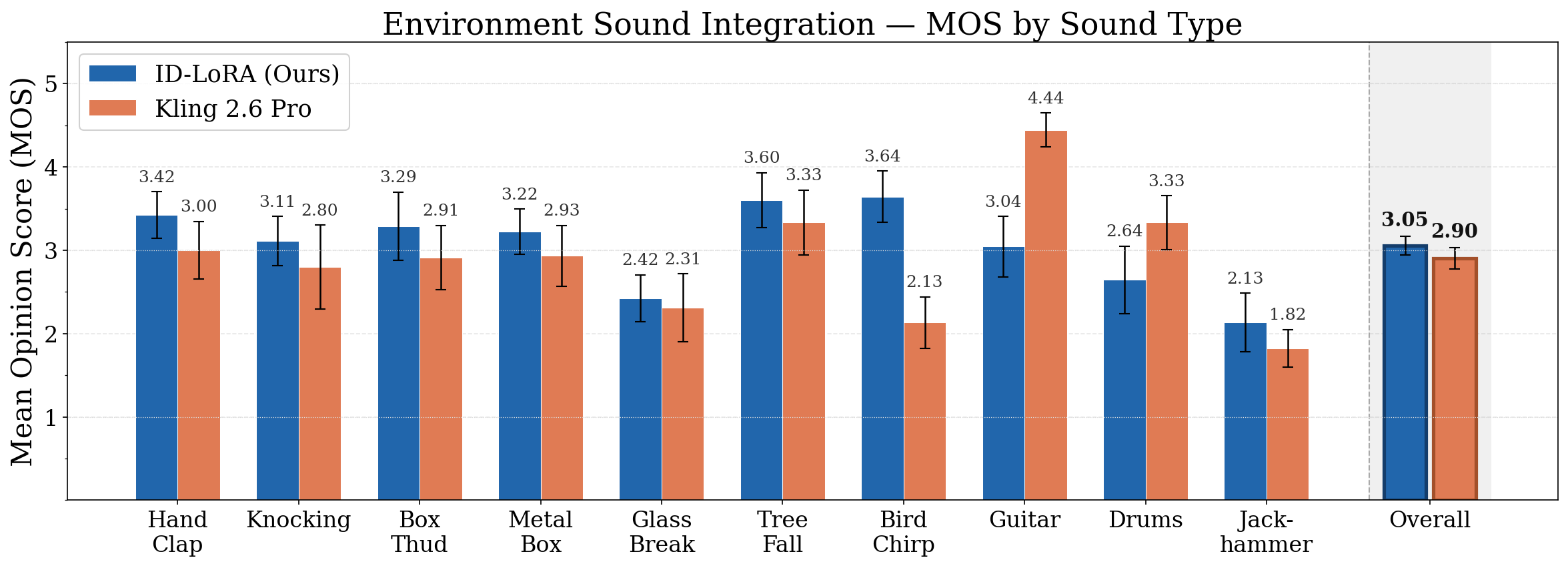}
\caption{Environment sound interaction MOS study: per-scenario mean opinion scores (1--5) with 95\% confidence intervals for \ourmethod and Kling 2.6 Pro across ten physical interaction scenarios. \ourmethod scores higher on 8 of 10 scenarios and in overall score.}
\label{fig:mos_env_interaction}
\end{figure}

  To complement automatic metrics, we conduct two human evaluations on
  Amazon Mechanical Turk (AMT), with 9 annotators per item. Participation is restricted to workers with Masters qualification, a HIT approval rate
  above 97\%, and location in English-speaking countries.

  \paragraph{A/B Preference Study.}
  We perform two A/B preference tests comparing \ourmethod against
  Kling 2.6 Pro and ElevenLabs + WAN2.2, using the CelebV-HQ hard split. For each pair, annotators are shown videos from both methods in randomized A/B order and select a preference (A, B, or Equal) along three axes: voice similarity (with a reference video shown), environment sounds (with the target description shown), and speech manners (with the target style description shown).
  To ensure annotation quality, we employ A/B randomization independent
  per (pair, question), a cyclic Latin square so that no annotator
  evaluates the same pair under two different question types, watch-gating
  that disables rating until all videos are watched to completion, guided
  practice rounds, and two hidden control pairs with known answers to
  filter inattentive workers (see supplementary for full protocol).

  Figure~\ref{fig:user_study} shows the preference rates, with
  ${\sim}285$--$290$ annotations per question type. All p-values are
  from two-sided binomial tests on win counts excluding equal responses.
  Against ElevenLabs + Wan 2.2, \ourmethod is overwhelmingly preferred
  for voice similarity (80.7\% vs.\ 17.5\%; win rate 82.1\%,
  $p < 0.001$) and environment sounds (68.7\% vs.\ 5.6\%; win rate
  92.4\%, $p < 0.001$), with a moderate advantage for speech manners
  (55.5\% vs.\ 39.9\%; win rate 58.1\%, $p < 0.01$).
  Against Kling 2.6 Pro, \ourmethod is significantly preferred on all
  three axes ($p < 0.001$): voice similarity (73.1\% vs.\ 20.0\%;
  win rate 78.5\%), environment sounds (54.8\% vs.\ 20.7\%; win rate
  72.6\%), and speech manners (65.2\% vs.\ 30.7\%; win rate 68.0\%).
  Across both comparisons, the strongest advantages appear in voice
  similarity and environment sounds, validating that unified generation
  jointly preserves speaker identity and enables text-driven control
  over the audio scene.

  \paragraph{Environment Sound Interaction MOS Study.}
  The A/B study evaluates ambient sounds described in the prompt
  (\eg, wind, traffic). A more demanding test is whether the model can
  generate sounds corresponding to \textit{physical interactions}
  depicted in the scene---\eg, a thud when a box drops or guitar music
  when strumming. Cascaded baselines cannot perform this task as they
  generate speech only; we therefore compare \ourmethod against
  Kling 2.6 Pro only.
  We design ten interaction scenarios, \eg dropping a box, clapping, drumming. For each scenario, we select five speakers
  from TalkVid and use Nano-banana~\cite{nanobanana_pro_2026} to insert a scene-appropriate object and background into the first frame (\eg, a guitar in the
  speaker's hands). Both methods generate videos from these edited frames
  with prompts describing the interaction, yielding new 50 paired samples.
  Annotators rate each video independently on a 1--5 MOS scale
  (1\,=\,Bad, 5\,=\,Excellent) assessing how well the generated audio
  matches the depicted interaction.
  Representative examples are shown in Fig.~\ref{fig:mos_examples};
  Scenario details and per-scenario analysis are in the supplementary.

  Results are shown in Figure~\ref{fig:mos_env_interaction}. \ourmethod
  achieves a higher overall MOS (3.05 vs.\ 2.90), wins on 8 of 10
  scenarios, and exhibits lower variance with fewer ``Bad'' ratings
  (15.3\% vs.\ 23.3\%). The fact that \ourmethod, trained on only ${\sim}$3K pairs, matches a large-scale commercial system confirms that unified generation provides a strong inductive bias for physically grounded
  audio-visual correspondence.

\section{Conclusions}
\label{sec:conclusions}

We presented \textbf{\ourmethod}, a unified audio-visual personalization method that jointly synthesizes a subject's appearance and vocal identity. Unlike cascaded pipelines, our approach benefits from cross-modal attention and allows text conditioning to influence both modalities simultaneously by leveraging the LTX-2 backbone for zero-shot personalization. Key architectural contributions include negative temporal positions for reference-target separation and identity guidance for enhanced speaker preservation.

Experiments demonstrate that \ourmethod outperforms cascaded baselines and competes with state-of-the-art closed-source models in speaker similarity and lip synchronization. Crucially, human evaluations show our method significantly surpasses both cascaded baselines and Kling 2.6 Pro in matching speech manners, generating appropriate environment sounds, and integrating physically grounded audio. Overall, \ourmethod yields more coherent audio-visual outputs with stronger prompt adherence and identity preservation than dedicated TTS pipelines. Future work will explore multi-speaker, cross-lingual, and general Audio-Visual Reference-to-Video generation.

\newpage
\section*{Acknowledgments}
We thank the LTX team at Lightricks for their valuable support throughout this project.

\bibliographystyle{plainnat}
\bibliography{main}

\newpage

\appendix
\section*{Appendix}

\noindent This Appendix provides additional details and results that complement the main paper.

We organize the content as follows:
\begin{itemize}
    \item \textbf{Section~\ref{sec:impl_details}}: Additional implementation details.
    \item \textbf{Section~\ref{sec:additional_results}}: Additional quantitative results.
    \item \textbf{Section~\ref{sec:additional_qualitative}}: Additional qualitative results.
    \item \textbf{Section~\ref{sec:user_study_details}}: Human evaluation details.
    \item \textbf{Section~\ref{sec:ethics}}: Broader impact and ethics.
\end{itemize}

\section{Additional Implementation Details}
\label{sec:impl_details}

\subsection{Dataset Details}
\label{sec:supp_datasets}

\paragraph{Training Data.} We train \ourmethod on two datasets with separate checkpoints:

\begin{enumerate}
    \item \textbf{CelebV-HQ}~\cite{zhu2022celebv}: a large-scale high-quality video dataset containing diverse talking-head videos with varied acoustic environments and speaking styles.
    \item \textbf{TalkVid}~\cite{chen2025talkvid}: a large-scale dataset of upper-body talking-head videos spanning 7,729 speakers across 15+ languages, featuring high-resolution footage (1080p--4K) curated through a multi-stage pipeline. TalkVid exhibits less acoustic diversity than CelebV-HQ; its videos are predominantly recorded in controlled settings with fewer background sounds and less environmental variation.
\end{enumerate}

Both datasets are preprocessed through the following pipeline:
\begin{enumerate}
    \item Filtering videos without audio or with insufficient speech;
    \item Trimming leading and trailing silence;
    \item Segmenting into 121-frame clips at 25fps ($\sim$4.8 seconds);
    \item Generating video captions (visual and speech) and audio-only captions (speaking style and environmental sounds) via Gemini;
    \item Retaining only English-language samples;
    \item Clustering speakers using face embeddings;
    \item Retaining only speakers with $\geq 2$ segments.
\end{enumerate}

\paragraph{Pair Construction.}  We then construct reference-target pairs from each speaker's clip pool. For CelebV-HQ, we balance the training set to contain 50\% same-video pairs and 50\% cross-video pairs, encouraging the model to generalize across both settings, and filter out pairs whose speaker similarity falls below 0.45 (measured via WavLM+ECAPA-TDNN), ensuring that the model trains only on pairs with reliable identity correspondence. For TalkVid, we form same-speaker reference-target pairs from the resulting 752 speaker clusters (7{,}862 segments). We apply Silero VAD filtering to remove segments where source separation inadvertently stripped speech. 
For both datasets, we then partition speakers into disjoint train and test sets (approximately 80/20 split), ensuring no speaker identity leakage between splits.
The resulting training set comprises approximately 3{,}000 pairs across 1{,}295 speakers for CelebV-HQ and 5{,}796 pairs across 600 speakers for TalkVid.

\paragraph{Source Separation.} Critically, we apply source separation to strip background sounds from all reference audio clips, providing clean speech-only references during training. This prevents the model from simply copying the reference environment into the output and instead forces it to rely on the text prompt for environment sounds and speaking style, which is essential for prompt-adherent audio generation.

\paragraph{Evaluation Splits.}
For CelebV-HQ, we curate two evaluation splits from a held-out set of speakers not seen during training:
\begin{itemize}
    \item \textit{Easy} (same-video): reference and target clips come from different segments of the same source video. This setting represents voice replication with high source-target speaker similarity.
    \item \textit{Hard} (cross-video): reference and target clips come from different videos of the same speaker. This tests generalization to new videos where acoustic conditions and speaking style vary significantly between source and target, resulting in lower baseline speaker similarity.
\end{itemize}

\noindent
Overall, we curate 35 video pairs in the hard split, and 36 video pairs in the easy split.

For TalkVid, we curate a quiet-to-environment benchmark from held-out speakers not seen during training. Each pair consists of a quiet (clean) reference clip and a target clip containing environmental sounds, testing whether the model can generate appropriate environment sounds from a clean reference while preserving speaker identity. The final set contains 49 pairs across 41 speakers (18 same-video, 31 cross-video).

Unlike the train set, the test splits did not go through source separation to keep the evaluation dataset similar to real-world data distribution.

\subsection{Baseline Details}
\label{sec:supp_baselines}

We compare \ourmethod against three cascaded pipelines and a unified commercial model. We note that Audiobox~\cite{vyas2023audiobox}, while the closest audio-only work to ours, is closed source and currently unavailable, precluding direct comparison.

\paragraph{Cascaded Baselines.} All three cascaded baselines follow a two-stage pattern: a zero-shot voice-cloning model generates speech from a reference audio clip and text transcript, then a sound-conditioned image-to-video model (WAN2.2~\cite{wan2025wanopenadvancedlargescale}) produces the final video from this audio and a first-frame image. We evaluate three TTS models:
\begin{itemize}
    \item \textbf{CosyVoice 3.0}~\cite{du2024cosyvoice}: a flow-based multilingual voice cloning model.
    \item \textbf{VoiceCraft}~\cite{peng-etal-2024-voicecraft}: an autoregressive neural codec language model for zero-shot speech editing and TTS.
    \item \textbf{ElevenLabs}~\cite{elevenlabs2026}: a state-of-the-art commercial voice cloning platform with built-in emotion and style control. We applied their automatic prompt enhancement to adjust pacing, emphasis, and emotion following their official recommendations, making it the strongest cascaded baseline for speaking style adherence.
\end{itemize}

\paragraph{Kling 2.6 Pro.} Kling 2.6 Pro~\cite{kling2026} is a closed-source commercial model that generates talking-face video with voice cloning in a unified (non-cascaded) pipeline. It represents a strong industry baseline for audio-visual personalization.

\subsection{Evaluation Metric Details}
\label{sec:supp_metrics}

\paragraph{Speaker Similarity.} We measure speaker identity preservation using WavLM Large~\cite{chen2022wavlm} combined with ECAPA-TDNN~\cite{desplanques2020ecapa}, following the UniSpeech-SAT speaker verification framework~\cite{chen2022unispeech}. We report cosine similarity between generated and target audio embeddings.

\paragraph{Face Similarity.} Visual identity is measured using ArcFace~\cite{deng2019arcface} embeddings extracted from detected faces. We average embeddings across 5 uniformly sampled frames and report cosine similarity to the target face.

\paragraph{Lip Synchronization.} Audio-visual alignment is evaluated using SyncNet~\cite{chung2016out} metrics: LSE-D (Lip Sync Error - Distance, lower is better) and LSE-C (Lip Sync Error - Confidence, higher is better). These metrics quantify how well lip movements match the generated audio.

\paragraph{Audio Prompt Adherence.} To evaluate how faithfully the generated audio follows the text prompt's intended environment and speaking style, we compute CLAP~\cite{elizalde2023clap} similarity between the generated audio and the combined prompt description of environment sounds and speaking style. Higher CLAP scores indicate better adherence to the prompt-specified audio properties. In cascaded pipelines, the voice-cloning stage tends to override these prompt-specified properties with the reference clip's acoustics; our unified approach enables the text prompt to influence the generated audio directly.

\paragraph{Word Error Rate (WER).} We measure speech intelligibility by transcribing the generated audio using Whisper-large-v3 and computing WER against the ground-truth transcript. Lower WER indicates clearer, more intelligible speech generation.

\section{Additional Quantitative Results}
\label{sec:additional_results}

\subsection{Face Similarity and Lip Motion Analysis}
\label{sec:supp_facesim}

All baselines achieve comparable or higher face similarity than \ourmethod, while Kling 2.6 Pro, another unified model, scores even lower (0.847--0.854).
To investigate whether this gap stems from reduced facial dynamics rather than superior identity preservation, we measure lip motion by extracting InsightFace~\cite{deng2019arcface} 68-point landmarks on every frame, computing the per-frame normalized inner mouth opening, and reporting its standard deviation over time.
As shown in Table~\ref{tab:facesim_lip_motion}, focusing on the top 30\% of samples by face similarity for each LTX method, our method consistently exhibits the highest lip motion across all three benchmarks while achieving the lowest FaceSim, and conversely LTX-Zeroshot, which produces no speech-driven animation, achieves the lowest lip motion yet the highest FaceSim by a wide margin (0.927 vs.\ 0.832 on TalkVid).
Qualitatively, we observe that the LTX-based cascaded pipelines frequently produce near-static frames with little to no facial movement, which trivially preserves the reference identity.
This reveals a systematic bias in ArcFace-based face similarity: less facial deformation keeps the face embedding closer to the target identity, inflating the metric independently of actual identity preservation quality. Face similarity alone is therefore an incomplete measure for talking-head video generation, as it inherently penalizes methods that produce more realistic speech articulation.

\begin{table}[t]
\centering
\caption{Lip motion ($\times 10^{2}$) vs.\ face similarity (FaceSim) for the top 30\% highest-FaceSim samples of each method. Higher lip motion indicates more active mouth articulation. Across all datasets, methods with less lip movement consistently achieve higher FaceSim, revealing a systematic metric bias against realistic speech animation.}
\label{tab:facesim_lip_motion}
\resizebox{\linewidth}{!}{%
\begin{tabular}{l cc cc cc}
\toprule
 & \multicolumn{2}{c}{\textbf{CelebVHQ-Hard}} & \multicolumn{2}{c}{\textbf{CelebVHQ-Easy}} & \multicolumn{2}{c}{\textbf{TalkVid}} \\
\cmidrule(lr){2-3} \cmidrule(lr){4-5} \cmidrule(lr){6-7}
\textbf{Method} & LipMot.$\uparrow$ & FaceSim$\uparrow$ & LipMot.$\uparrow$ & FaceSim$\uparrow$ & LipMot.$\uparrow$ & FaceSim$\uparrow$ \\
\midrule
LTX-Zeroshot      & 0.95 & \textbf{0.947} & 1.73 & \textbf{0.944} & 2.41 & \textbf{0.927} \\
VoiceCraft-MFA-LTX & 1.03 & 0.944 & 0.97 & 0.947 & 2.94 & 0.832 \\
CosyVoice-LTX     & 1.07 & 0.944 & 1.90 & 0.940 & 2.95 & 0.828 \\
ElevenLabs-LTX    & 1.29 & 0.943 & 1.29 & 0.942 & 2.82 & 0.829 \\
\midrule
Ours              & \textbf{1.56} & 0.936 & \textbf{2.98} & 0.939 & \textbf{3.07} & 0.832 \\
\bottomrule
\end{tabular}%
}
\end{table}

\subsection{Identity-guidance sensitivity sweep}

\paragraph{Identity Guidance Scale.} Table~\ref{tab:idg_ablation} analyzes the effect of the identity guidance scale $s_{\text{id}}$ on the hard split. At the lowest tested scale ($s_{\text{id}} = 2$), speaker similarity drops substantially (0.459 vs.\ 0.477 at $s_{\text{id}} = 4$), confirming that identity guidance is critical for voice preservation. Increasing the scale beyond the default ($s_{\text{id}} = 4$) further improves speaker similarity up to $s_{\text{id}} = 10$ (0.485), but with degrading lip synchronization and speech intelligibility at high scales. We select $s_{\text{id}} = 4$ as the default, balancing speaker identity with generation quality.

\begin{table}[t]
    \centering
    \caption{Identity-guidance scale sweep $s_{\text{id}}$ on the hard split (two-stage), with negative temporal positions enabled. Larger $s_{\text{id}}$ generally improves speaker similarity but introduces trade-offs in lip-sync and ASR metrics.}
    \label{tab:idg_ablation}
    \begin{tabular}{@{}lccccc@{}}
    \toprule
    $s_{\text{id}}$ & Spk Sim $\uparrow$ & Face Sim $\uparrow$ & LSE-D $\downarrow$ & CLAP $\uparrow$ & WER $\downarrow$ \\
    \midrule
    2 & 0.459 & 0.872 & 8.63 & 0.348 & 0.117 \\
    \textbf{4 (default)} & 0.477 & 0.874 & 8.50 & \textbf{0.363} & \textbf{0.113} \\
    6 & 0.477 & 0.874 & \textbf{8.45} & 0.355 & 0.142 \\
    8 & 0.481 & 0.873 & 8.52 & 0.361 & 0.185 \\
    10 & \textbf{0.485} & 0.872 & 8.56 & 0.353 & 0.119 \\
    12 & 0.482 & \textbf{0.876} & 8.69 & 0.350 & 0.145 \\
    14 & 0.473 & 0.872 & 8.73 & 0.358 & 0.147 \\
    16 & 0.470 & 0.872 & 8.89 & 0.358 & 0.143 \\
    \bottomrule
    \end{tabular}
    \end{table}

\section{Additional Qualitative Results}
\label{sec:additional_qualitative}

Please refer to our website \url{https://id-lora.github.io/} for high-resolution video comparisons and qualitative examples

\section{Human Evaluation Details}
\label{sec:user_study_details}

\subsection{A/B Preference Study Protocol}
\label{sec:supp_ab_protocol}

We conduct two separate A/B preference tests on Amazon Mechanical Turk (AMT), each comparing \ourmethod against one baseline: Kling 2.6 Pro and ElevenLabs + WAN2.2. All studies use the \textit{hard} (cross-video) evaluation split comprising 35 video pairs across 8 held-out CelebV-HQ speakers. Each pair is evaluated by 9 annotators. We restrict participation to workers with Masters qualification, a HIT approval rate above 97\%, and location in English-speaking countries (US, UK, Canada, Australia).

For each video pair, annotators are shown videos generated by \ourmethod and the baseline in randomized A/B order and asked to select a preference (A, B, or Equal) along three axes:

\begin{enumerate}
    \item \textbf{Voice Similarity}: ``Which video's voice sounds more like the reference?'' (a reference video of the target speaker is shown alongside the two candidates).
    \item \textbf{Environment Sounds}: ``Which video's background sounds better match the description?'' (the target environment sounds description is displayed).
    \item \textbf{Speech Manners}: ``Which video's speaking style better matches the description?'' (the target speaking style description is displayed).
\end{enumerate}

\begin{figure}[t]
    \centering
    \begin{subfigure}[t]{0.56\linewidth}
        \centering
        \includegraphics[width=\linewidth]{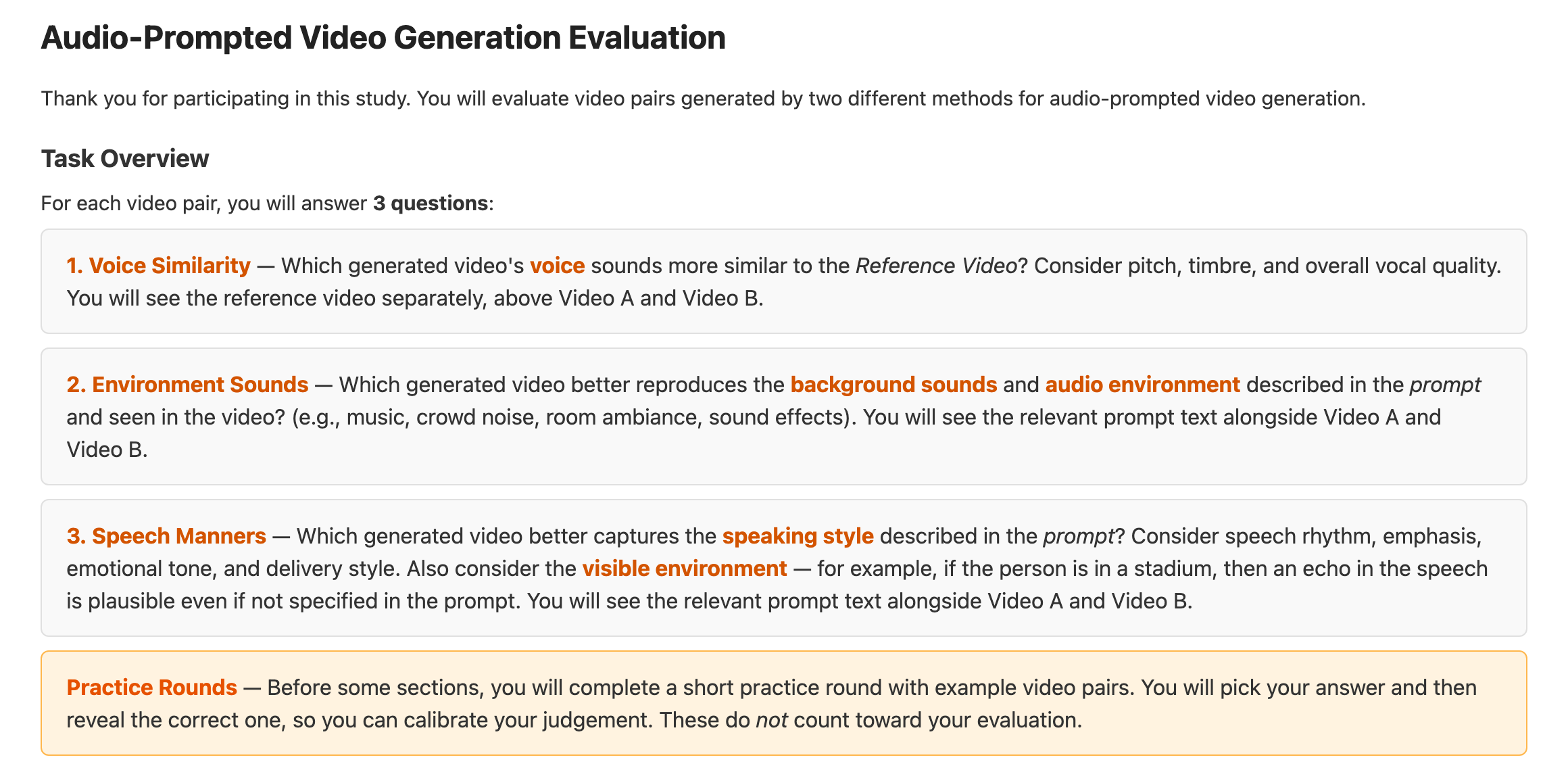}
        \caption{A/B preference study: annotators are briefed on the three evaluation axes and the practice round format.}
        \label{fig:ab_instructions}
    \end{subfigure}
    \hfill
    \begin{subfigure}[t]{0.41\linewidth}
        \centering
        \includegraphics[width=\linewidth]{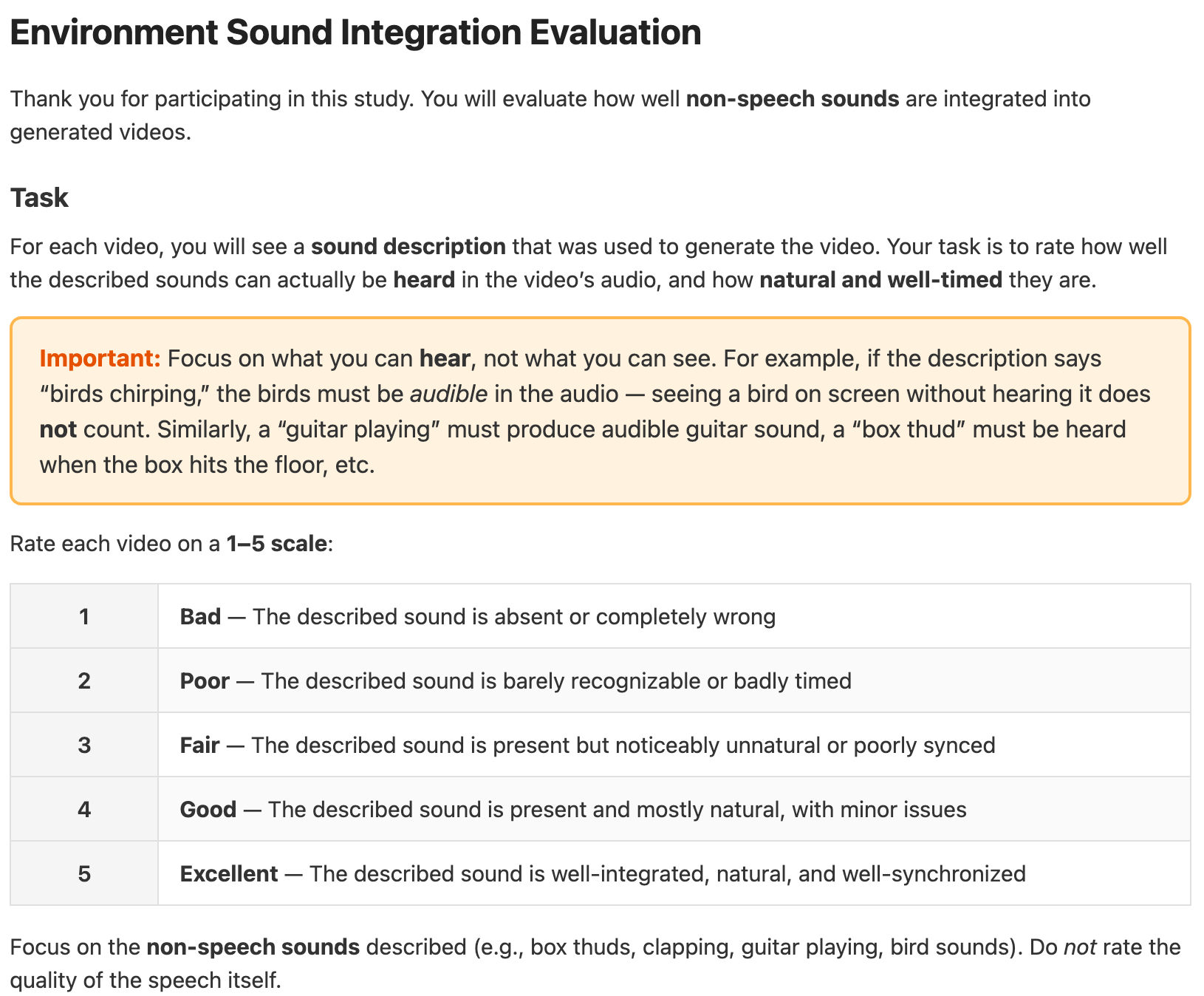}
        \caption{MOS study: annotators rate how well the described non-speech sounds can be heard on a 1--5 scale.}
        \label{fig:mos_instructions}
    \end{subfigure}
    \caption{Instruction pages shown to Amazon Mechanical Turk annotators at the beginning of each human evaluation study.}
    \label{fig:study_instructions}
\end{figure}

\noindent The instruction pages for both studies are shown in Figure~\ref{fig:study_instructions}, and example evaluation cards are shown in Figure~\ref{fig:study_examples}.

\paragraph{Bias Prevention.} We employ several mechanisms to ensure annotation quality:
\begin{itemize}
    \item \textbf{A/B randomization}: The A/B assignment is randomized independently per (pair, question), so the same pair may have the methods in opposite positions for different questions.
    \item \textbf{Cyclic Latin square}: A cyclic Latin square distributes pairs across annotators such that no annotator evaluates the same pair for two different question types, preventing familiarity bias.
    \item \textbf{Watch-gating}: Rating buttons are disabled until all videos have been watched to completion; forward-scrubbing is blocked on unanswered pairs.
    \item \textbf{Practice rounds}: Guided examples with ``Reveal Answer'' feedback are provided before the environment sounds and speech manners sections to calibrate annotators.
    \item \textbf{Hidden control pairs}: Two hidden control pairs with known correct answers (one for environment sounds, one for speech manners) are embedded in the evaluation to identify inattentive annotators. Workers who fail any control are excluded from analysis.
\end{itemize}

\subsection{Environment Sound Interaction MOS Study}
\label{sec:supp_mos}

The A/B preference study evaluates ambient environment sounds described in the text prompt (\eg, wind, music, traffic). A more demanding test of audio prompt adherence is whether the model can generate sounds corresponding to \textit{physical interactions} described in the prompt and depicted in the scene for example, producing a thud when the prompt describes a box dropping, or guitar music when the prompt describes strumming.

\paragraph{Why Only Unified Models.} Cascaded voice-cloning baselines \textit{cannot} perform this task: they generate clean speech only and have no mechanism to produce non-speech sounds. More broadly, any audio-only model generating without access to the evolving visual scene cannot accurately time acoustic events to on-screen actions: if a bird is already mid-flight in the first frame, a blind model has no visual cues about when the interaction begins or ends. Even Audiobox~\cite{vyas2023audiobox}, which combines voice cloning with text-described background sounds, operates purely in the audio domain and cannot condition on visual context. This evaluation is therefore restricted to unified audio-visual models. We compare \ourmethod against Kling 2.6 Pro as the state-of-the-art commercial unified model.

\begin{figure}[t]
    \centering
    \begin{subfigure}[t]{0.48\linewidth}
        \centering
        \includegraphics[width=\linewidth]{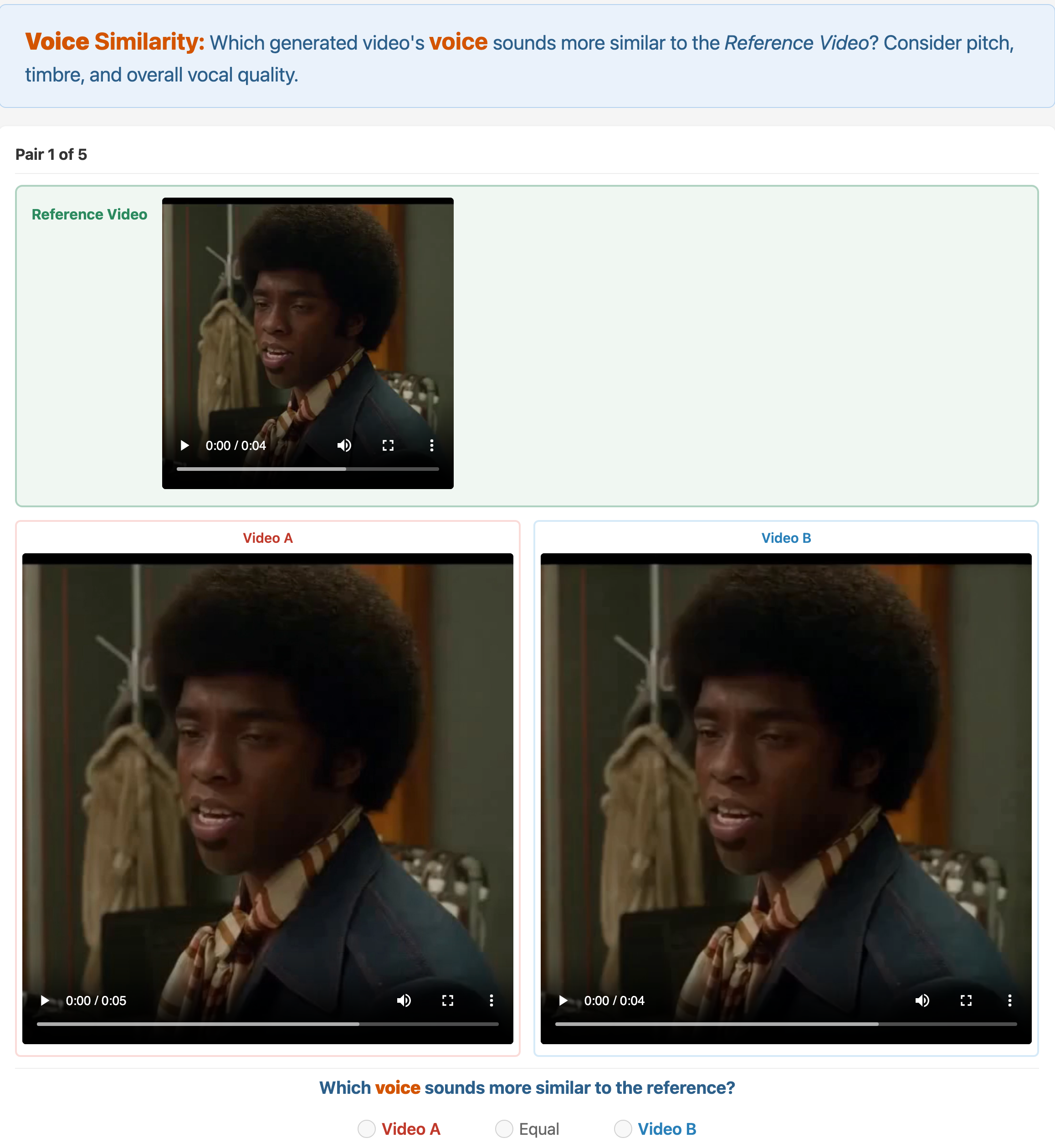}
        \caption{A/B preference study (voice similarity). Annotators watch a reference video alongside two generated videos and select which voice sounds more similar, or equal.}
        \label{fig:ab_speech}
    \end{subfigure}
    \hfill
    \begin{subfigure}[t]{0.48\linewidth}
        \centering
        \includegraphics[width=\linewidth]{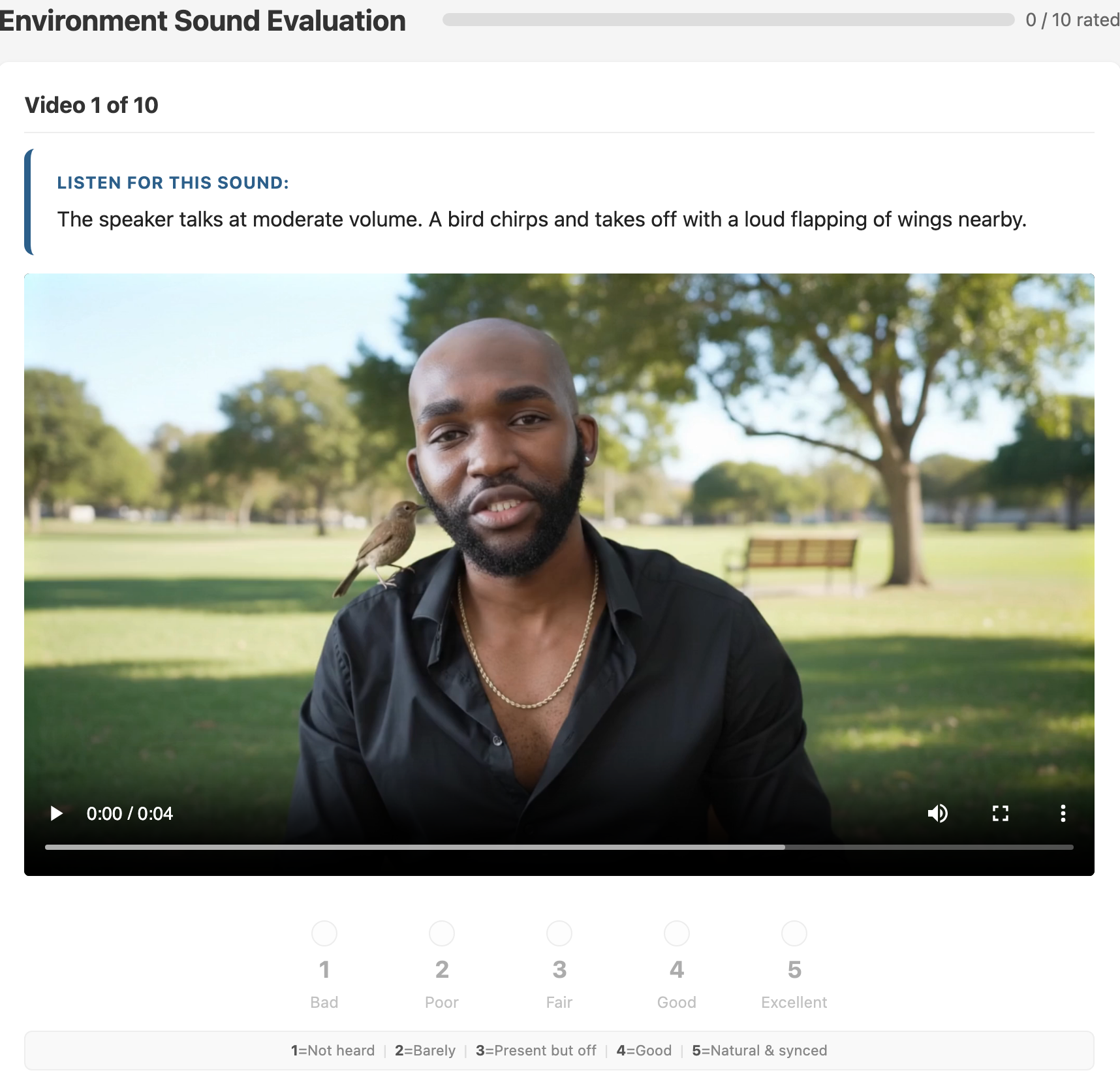}
        \caption{MOS study (bird chirping scenario). Annotators see the sound description, watch the video, and rate the sound quality on a 1--5 scale.}
        \label{fig:mos_card}
    \end{subfigure}
    \caption{Example evaluation cards from both human evaluation studies.}
    \label{fig:study_examples}
\end{figure}

\paragraph{Scenario Design.} We design a MOS study covering ten interaction scenarios: \textit{box dropping, clapping, drumming, glass breaking, guitar playing, jackhammer, door knocking, metal box hitting, bird chirping in a park}, and \textit{tree falling}. For each scenario, we select five random speakers from the TalkVid dataset and use Nano Banana Pro~\cite{nanobanana_pro_2026} to edit the first frame of each speaker's video to insert a scene-appropriate object (\eg, a guitar in the speaker's hands, a person working with a jackhammer in the background). Both methods then generate videos from these edited first frames with text prompts describing the corresponding interaction sounds, yielding 50 paired comparisons. We collect 450 ratings per method from AMT annotators (9 annotators per video) on a 1--5 MOS scale (1 = Bad, 5 = Excellent). The instruction page is shown in Figure~\ref{fig:mos_instructions}, and an example evaluation card is shown in Figure~\ref{fig:mos_card}.

\paragraph{Per-Scenario Analysis.} \ourmethod achieves a higher overall MOS than Kling 2.6 Pro (3.05 vs.\ 2.90) and wins on 8 of 10 scenarios, with the largest gains on bird chirping in a park (+1.51), clapping (+0.42), and box dropping (+0.38). Kling outperforms on guitar ($-1.40$) and drums ($-0.69$), both sustained musical instrument sounds where its larger-scale training likely provides an advantage. \ourmethod also exhibits lower variance and a smaller proportion of ``Bad'' ratings (15.3\% vs.\ 23.3\%), suggesting more consistent generation quality.

While the overall difference does not reach statistical significance ($p = 0.093$, Cohen's $d = 0.115$), the fact that \ourmethod, a parameter-efficient adaptation trained on ${\sim}$3K pairs, matches a large-scale commercial system confirms that unified generation provides a powerful inductive bias for physically grounded audio-visual correspondence.

We view this study as an exploratory first step toward evaluating diegetic consistency in unified audio-visual generation. By combining first-frame editing with unified generation, this protocol opens new directions for benchmarking physically grounded audio-visual correspondence, for example, testing whether reverberation adapts to scene geometry or whether impact sounds respect material properties, capabilities that are uniquely enabled by unified architectures.

\section{Broader Impact and Ethics}
\label{sec:ethics}

\paragraph{Potential Risks.} \ourmethod generates realistic audio-visual content that preserves a specific person's face and voice from a short reference clip. This capability, while valuable for legitimate applications, carries risks of misuse. Bad actors could produce non-consensual impersonations, fabricate misleading media, or clone a person's likeness without authorization.

\paragraph{Mitigations.} We advocate for several safeguards to reduce misuse potential. First, generated content should carry invisible watermarks or provenance metadata (\eg, C2PA) to enable downstream detection and attribution. Second, deployment should require explicit consent from depicted individuals, particularly for voice cloning, where legal frameworks such as the EU AI Act increasingly mandate transparency and consent. Third, continued investment in deepfake detection research is essential to keep pace with generation advances. We note that \ourmethod does not introduce fundamentally new risks beyond those already posed by existing commercial systems (Kling, ElevenLabs, HeyGen) that offer similar capabilities at scale; rather, it provides an open research framework for studying and mitigating these risks.

\paragraph{Positive Applications.} Unified audio-visual personalization enables several beneficial use cases: multilingual dubbing that preserves a speaker's identity and matches scene acoustics, accessibility tools that generate personalized videos, digital avatars for individuals with speech impairments, and creative content production where actors can be depicted in scenes that would be impractical or unsafe to film.

\paragraph{Data and Human Evaluation Ethics.} Our training datasets (CelebV-HQ and TalkVid) are publicly available research datasets. Our human evaluation studies on Amazon Mechanical Turk followed standard ethical practices: annotators were compensated above minimum wage, participation was voluntary, and no personally identifiable information was collected beyond anonymized worker IDs. The study was designed to minimize annotator burden through watch-gating and practice rounds.


\end{document}